\documentclass[usenatbib]{mn2e}
\usepackage[utf8]{inputenc}
\usepackage{natbib,epsfig,cite,mystyle,amsmath,amssymb,graphicx}
\usepackage{array, xcolor, caption}
\usepackage{hyperref}
\hypersetup{colorlinks,allcolors=black}

\pdfoutput=1  % togliere il commento se le figure sono pdf, lasciarlo se sono eps
\usepackage{bm}
\usepackage{graphicx,color}
\usepackage{latexsym,amsmath,amssymb,graphicx,booktabs}
\usepackage{placeins}
\usepackage{subfigure}
\usepackage{mathtools}
\usepackage[normalem]{ulem}

\definecolor{MyBlue}{rgb}{0.15,0.15,0.70}
\definecolor{Dgreen}{rgb}{0,0.7,0.0}

\hypersetup{
colorlinks=true,
citecolor=MyBlue,
linkcolor=MyBlue,
urlcolor=MyBlue
}

\usepackage{amssymb}
\usepackage{amsmath}
\usepackage{amsfonts}
\usepackage{upgreek}
\usepackage{latexsym}
\usepackage{appendix}
\usepackage{dsfont}
\usepackage{mathabx}

\newcommand\spart{\;\raise1.0pt\hbox{/}\hskip-6pt\partial}
\newcommand\spartb{\;\overline{\raise1.0pt\hbox{/}\hskip-6pt\partial}}

\newcommand{\be}{\begin{equation}}
\newcommand{\ee}{\end{equation}}

\newcommand{\beqa}{\begin{eqnarray}}
\newcommand{\eeqa}{\end{eqnarray}}

\begin{document}
\title[Noisy neighbours]{Noisy neighbours: inference biases from overlapping gravitational-wave signals}
\author[A.~Antonelli, O.~Burke, J.R.~Gair]{Andrea Antonelli$^{1}$\thanks{andrea.antonelli@aei.mpg.de}, Ollie Burke$^{1,2}$\thanks{ollie.burke@aei.mpg.de}, Jonathan R.~Gair$^{1,2}$\thanks{jonathan.gair@aei.mpg.de} \\
$^{1}$Max Planck Institute for Gravitational Physics (Albert Einstein Institute), Am M\"{u}hlenberg 1, Potsdam-Golm, 14476, Germany \\
$^{2}$School of Mathematics, University of Edinburgh, James Clerk Maxwell Building, Peter Guthrie Tait Road, Edinburgh EH9 3FD, UK
}

\pagerange{\pageref{firstpage}--\pageref{lastpage}} \pubyear{2021}
\maketitle
\label{firstpage}

\begin{abstract}
Understanding and dealing with inference biases in gravitational-wave (GW) parameter estimation when a plethora of signals are present in the data is one of the key challenges for the analysis of data from future GW detectors. Working within the linear signal approximation, %Fisher formalism, 
we describe generic metrics to predict inference biases on GW source parameters in the presence of confusion noise from unfitted foregrounds, from overlapping signals that coalesce close in time to one another,
and from residuals of other signals that have been incorrectly fitted out. We illustrate the formalism with simplified, yet realistic, scenarios appropriate to third-generation ground-based (Einstein Telescope) and space-based (LISA) detectors, and demonstrate its validity against Monte-Carlo simulations. We find it to be a reliable tool to cheaply predict the extent and direction of the biases. 
Finally, we show how this formalism can be used to correct for biases that arise in the sequential characterisation of multiple sources in a single data set, which could be a valuable tool to use within a global-fit analysis pipeline.

\end{abstract}


%%%%%%%%%%%%%%%%%%%%%%%%%%%%%%%%%%%%%%%%%%%%%%%%%%
\section{Introduction} 
%%%%%%%%%%%%%%%%%%%%%%%%%%%%%%%%%%%%%%%%%%%%%%%%%%

In the analysis of data from future gravitational-wave (GW) detectors, we will be confronted with the prospect of detecting and performing parameter inference on sources that overlap with other resolved or unresolved signals. The presence of such additional signals in the data or their incomplete removal through inaccurate waveform templates, might lead to biases in the parameter estimates for the source of interest, if they are not properly accounted for. While this possibility is relevant for imminent upgrades of the LIGO-Virgo-KAGRA detectors' network~\citep{Aasi:2013wya}, the odds of this happening are higher with future ground-based and space-based detectors such as the Einstein Telescope (ET)\citep{Punturo:2010zz}, Cosmic Explorer (CE)~\citep{Reitze:2019iox} and the Laser Interferometer Space Antenna (LISA)\citep{Audley:2017drz}. 
The former is expected to detect thousands of GW signals from low-mass black holes and neutron stars~\citep{Punturo:2010zz}, the latter is guaranteed to detect tens of thousands of white dwarf binaries in the Milky Way, and is also expected to detect signals from mergers involving supermassive black holes ~\citep{Audley:2017drz}. For these future detectors, one will \emph{have to} take into account the possible presence of signals or high-SNR residuals lurking in the data. As this problem is only of peripheral relevance to analyses for the current LIGO-Virgo detector network, it has attracted relatively limited attention in the literature. In the context of ground-based detector networks, the detectability of confusion noise from a population of unresolved signals has been considered~\citep{Regimbau:2009rk}, but not the impact of the presence of that confusion foreground on parameter estimation for resolved sources. There have also been some recent Bayesian parameter estimation studies for second and third generation detectors, which computed the bias that arises in parameter estimation for a source due to the presence of another source with an overlapping merger~\citep{Samajdar:2021egv,Pizzati:2021gzd,Himemoto:2021ukb,Relton:2021cax}, and the impact of simultaneous fitting of two sources on the individual parameter precisions~\citep{1853026}. These studies were limited to just two sources and did not consider the impact of waveform modelling uncertainties. 
In the LISA context, there have been studies of the detectability of confusion foregrounds from unresolved extreme mass-ratio inspirals~\citep{Bonetti_2020}, and extensive exploration of the simultaneous global-fit of the thousands of galactic binary sources expected to be present in LISA data~\citep{Robson:2017ayy,Littenberg:2020bxy,Karnesis:2021tsh}. The latter global-fit analyses tackle the problem head-on by considering the simultaneous inference on parameters of an unknown number of sources in the data stream. Clearly, this is a formidable task due to the exceptionally large parameter space and complexity of the likelihood surface.
It is thus important to have independent procedures to
aid global-fit search pipelines (and potentially confirm the results).

We use semi-analytic methods based on the Fisher formalism
to cheaply assess when confusion from other sources, and/or imperfect subtraction of those sources due to waveform errors, is likely to be problematic, in the sense of leading to significant biases in parameter estimation for a source of interest. We leverage existing metrics for the ``goodness'' of individual waveform models based on the linear signal (Fisher matrix) formalism to derive generic metrics to assess the inference biases on source parameter characterisations. We describe how to apply this approach to several cases of relevance: i) parameter estimation in the presence of ``confusion noise'' from unfitted signals in the data; ii) parameter estimation for two overlapping signals with approximately coincident coalescence times; iii) parameter estimation for a population of sources using inaccurate waveform models; and iv) the case in which both confusion noise and mismodelling errors contribute to the final biases. Finally, we will show how these results can be used to mitigate biases in a sequential-fitting pipeline for LISA.
Our analysis is related to previous work by~\citep{Flanagan:1997kp}, \citep{PhysRevD.71.104016} and~\citep{Cutler:2007mi}, in which expressions are provided for the error on parameters due to the presence of noise and due to waveform errors. While their work has been mainly considered in the context of individual signals in the data, two observations make it relevant and easily extendible to the above applications. Firstly, no assumptions are made on the source of the noise
appearing in their expressions, meaning that the observed noise can be made into a linear combination of detector noise and confusion noise [with applications to points i) and ii)]. Secondly, no assumptions are made about the dimensions of the parameter space, meaning that expressions relevant to points iii) and iv) can be derived from them. 

We illustrate these metrics for several cases of relevance to future ground-based and space-based detectors. We take the ET and LISA instruments as our examples and use simplified, but realistic models for the gravitational waveforms. We consider the following, increasingly more complex, situations:
\begin{itemize}
    \item The parameter estimation of a single LISA massive black hole source in the presence of other unfitted massive black holes forming a foreground [Sec.~(\ref{sec:source_conf})].
    \item The parameter estimation of a single ET source in the presence of an overlapping signal with time of coalescence a fraction of a second from the former [Sec.~(\ref{subsec:overlap_merger})].
    \item The parameter estimation of a single LISA source in the presence of two overlapping sources which have been incorrectly fitted out of the data [Sec.~(\ref{subsec:biases_innacurate_removal})].
    \item The simultaneous inference in LISA of a few overlapping sources, subject to waveform errors, detector noise and unresolved signals present in the data stream [Sec.~(\ref{sec:results})].
\end{itemize}
We find that unfitted foregrounds or incorrectly removed sources may lead to significant biases [as discussed in sections~Sec.~(\ref{sec:source_conf}) and~Sec.~(\ref{subsec:biases_innacurate_removal})], but that biases from confusion noise and waveform inaccuracies could deconstructively interfere [as discussed in Sec.~(\ref{sec:results})]. We qualitatively confirm one of the main results of~\citep{Samajdar:2021egv,Pizzati:2021gzd,Himemoto:2021ukb,Relton:2021cax} in Sec.~(\ref{subsec:overlap_merger}), showing that biases arise when the difference between the coalescence times of two overlapping signals is smaller than a few tens of waveform periods, corresponding to a fraction of a second for the ground-based detector examples considered in those studies. We find that the formalism herein developed is capable of predicting the biases very well (as confirmed with MCMC analyses), which makes it a useful tool for exploratory studies of future detectors. 

Finally, in Section~\ref{sec:GF}, we introduce the \emph{local-fit strategy} as a possible approach to the global-fit in LISA data analysis.
This method separately fits the parameters of individual sources, and then uses the %\sout{computationally-inexpensive}
Fisher-based formalism presented in (\ref{sec:Source_Confusion_Bias}) to correct the biases that result in these estimates from ignoring the other sources in the data. 
Although a well-designed algorithm could, in principle, deliver a simultaneous fit to all sources for comparable computational cost, the local-fit algorithm is likely to be much easier to implement and to optimise.
We believe that this algorithm could therefore be used to aid global-fit strategies, for example by providing a quick estimate of the parameters of all sources, that could be used as a starting point for a simultaneous-fitting algorithm that then delivers the final joint posterior distribution.

The paper is organised as follows: Sec.~(\ref{sec:DA_concepts}) contains a review of basic data-analysis concepts needed throughout the paper; Sec.~(\ref{sec:Source_Confusion_Bias}) contains the description of the Fisher formalism herein developed; Sec.~(\ref{sec:model&noise}) contains a brief review of our choices of waveform models; Sec.~(\ref{sec:illustrations}) discusses the illustrations of the formalism described above; Sec~(\ref{sec:GF}) describes the local-fit strategy; and Sec.~(\ref{conclusions}) summarises our findings and describes some possible future avenues of investigation. In appendix~(\ref{app:geometry}), we discuss a geometrical interpretation for the errors from noise and the biases from mismodelling~\citep{Cutler:2007mi}; in appendix~(\ref{app:num_routines}) we describe the numerical methods used to  obtain the results reported in the previous sections; in appendix~(\ref{app:Fisher_Matrix}) we describe how we computed the Fisher matrices and how these were verified using MCMC analyses; finally, in appendix~(\ref{app:ET_gf}) we complement the LISA results of Sec.~\ref{sec:results} with results for ET. % of the present paper.

%%%%%%%%%%%%%%%%%%%%%%%%%%%%%%%%%%%%%%%%%%%%%%%%%%
\section{Data Analysis Concepts} \label{sec:DA_concepts}
%%%%%%%%%%%%%%%%%%%%%%%%%%%%%%%%%%%%%%%%%%%%%%%%%%
The data stream observed by a gravitational wave detector is a superposition of noise $n(t)$ intrinsic to the detector and a gravitational wave signal $h_{e}$ with ``true'' parameters $\boldsymbol{\theta}_{\text{tr}}$
\begin{equation}\label{eq:data}
    d(t) = h_{e}(t;\boldsymbol{\theta}_{\text{tr}}) + n(t).
\end{equation}
In general, the gravitational wave component is a combination of the signals from a number of individual sources. The consequences of this will be made explicit in Section~\ref{sec:Source_Confusion_Bias}. 
In this analysis, we make the usual assumption that the noise $n(t)$ is both stationary and Gaussian with zero mean. As a consequence of stationarity, the covariance of the noise in the frequency domain can be expressed by~\citep{wiener1930generalized,khintchine1934korrelationstheorie}
\begin{equation}\label{eq:Wiener-Khinchin-Theorem_freq}
    \langle \hat{n}(f)\hat{n}^{\star}(f')\rangle = \frac{1}{2}\delta(f - f')S_{n}(f).
\end{equation}
Here and throughout this paper, hatted quantities will denote the continuous time Fourier transform. In the above, $\delta$ denotes the Dirac delta function and $\langle \cdot \rangle$ denotes an ensemble averaging process. The quantity $S_{n}(f)$ denotes the (one-sided) power spectral density, which describes the distribution of power of the noise in the frequency domain. 

The ``loudness'' of a signal can be represented by the optimal matched filtering signal to noise ratio (SNR), the square of which is given by
\begin{equation}\label{eq:cont_SNR}
    \rho^{2} = (h|h) = 4\int_{0}^{\infty}\frac{|\hat{h}(f)|^{2}}{S_{n}(f)}df
\end{equation}
where we have defined the inner product for real valued time-series,
\begin{equation}\label{eq:inn_prod}
(a|b) = 4\text{Re}\int_{0}^{\infty}\frac{\hat{a}(f)\hat{b}^{\star}(f)}{S_{n}(f)}df.
\end{equation}
To make inference on parameters, one requires a probabilistic model on the data stream for given unknown parameters $\boldsymbol{\theta}$. As the noise $n(t)$ is stationary and Gaussian, the Whittle (log) likelihood can be used~\citep{whittle:1957} 
\begin{equation}\label{eq:whittle_likelihood}
    \log p(d|\boldsymbol{\theta}) \propto -\frac{1}{2}(d - h_{m}|d - h_{m}).
\end{equation}
We note that the gravitational wave component of the data stream in Eq.~\eqref{eq:data},  $h_{e}(t;\boldsymbol{\theta}_{\text{tr}})$, is the true signal which depends on parameters $\boldsymbol{\theta}_{\text{tr}}$ that we wish to infer. In \eqref{eq:whittle_likelihood}, we are denoting the signal by $h_{m}$, to allow for the possibility that there is a difference between the approximate waveform templates used to analyse the data, and the true signal, $h_{e}$, present in the data stream.

Finally, to quantify the precision of measurements on parameters, we will make use of the the linear signal approximation (LSA)~\citep{Finn:1992wt}. By considering a small perturbation $\boldsymbol{\theta} = \boldsymbol{\theta}_{\text{tr}} + \Delta \boldsymbol{\theta}$, one can expand the waveform model in the vicinity of the best-fit parameters as
\begin{equation}\label{eq:LSA}
    h_{m}(t;\boldsymbol{\theta}) \approx h_{m}(t;\boldsymbol{\theta}_{\text{tr}}) + \partial_{i}h_{m}(t;\boldsymbol{\theta}_{\text{tr}})\Delta \theta^{i},
\end{equation}
which is valid for $|\Delta \theta_{\text{bf}}^{i}|\ll 1$.% \sout{$|\Delta \theta_{\text{bf}}^{i}/\theta_{\text{bf}}^{i}|\ll 1$}. 
We are using the standard notation $\partial_{i} = \partial/\partial \theta^i$.
Substituting \eqref{eq:LSA} into \eqref{eq:whittle_likelihood} and restricting to the case that the model and true waveform agree, $h_{m} = h_{e}$ for all $\boldsymbol{\theta}$, one obtains
\begin{align}\label{eq:fishy_likelihood}
    -2\log p(d|\boldsymbol{\theta}) &= (\Delta \theta^{i} - \Delta \theta_{\text{noise}}^{i}) \Gamma_{ij}(\Delta \theta^{j} - \Delta \theta_{\text{noise}}^{j}). \\
     \Delta \theta_{\text{noise}}^{i} &= (\Gamma^{-1})^{ij}(\partial_{j}h|n)\,,
\end{align}
where $\Gamma_{ij}$ is the Fisher matrix, with components
\begin{equation}\label{eq:Fish_Matrix}
    \Gamma_{ij} = (\partial_{i}h|\partial_{j}h).
\end{equation}
In the derivation of Eq.\eqref{eq:fishy_likelihood}, we neglected higher order terms which scale like $\mathcal{O}(\rho^{-1})$. Thus this representation of the likelihood is only valid for high SNR. Notice that \eqref{eq:fishy_likelihood} is Gaussian and centered on $\theta^{i}_{\text{bf}} = \theta^{i}_{\text{tr}} + \Delta\theta^{i}_{\text{noise}}$. %, for $\theta^{i}_{\text{bf}}$ the choice of parameter value that maximizes \eqref{eq:fishy_likelihood} . 
Defining the statistic $\widehat{\Delta\theta^{i}} = \Delta\theta^{i}_{\text{noise}}$, one observes 
\begin{equation}
    \mathbb{E}[\widehat{\Delta\theta^{i}}] = 0, \quad \text{Cov}(\widehat{\Delta\theta^{i}},\widehat{\Delta\theta^{j}}) = (\Gamma^{-1})^{ij} + \mathcal{O}(\rho^{-1}).
\end{equation}
This implies that the statistic $\widehat{\Delta\theta^{i}}$ is unbiased with co-variance equal to the inverse of the Fisher matrix. In other words, the shift in the peak of the likelihood due to noise fluctuations is consistent with its width. 

In the derivation of \eqref{eq:fishy_likelihood}, we assumed the model template was consistent with the true gravitational waveform in the data set. We can relax this assumption and now consider $h_{e}\neq h_{m}$, 
%for all parameter values $\boldsymbol{\theta}$,
which leads to a mismodelling error $\delta h(\boldsymbol{\theta})= h_{e}(t;\boldsymbol{\theta}) - h_{m}(t;\boldsymbol{\theta})$. %Frequentist inference on $\boldsymbol{\theta}_{\text{tr}}$ focuses on maximizing the 
The maximum of the likelihood function is at a set of parameter values $\boldsymbol{\theta}_{\text{bf}}$ that are a solution to
\begin{equation}
    (\partial_{i}h_{m}(t;\boldsymbol{\theta}_{\text{bf}})|d - h_{m}(t;\boldsymbol{\theta}_{\text{bf}})) = 0.
\end{equation}
Using the LSA \eqref{eq:LSA} and considering a perturbation $\boldsymbol{\theta}_{\text{tr}} = \boldsymbol{\theta}_{\text{bf}} + \Delta \boldsymbol{\theta}$ and a data stream $d(t) = h_e(t;\boldsymbol{\theta}_\text{tr})+n(t)$ including the true gravitational waveform, one obtains
\begin{align}\label{eq:CV_1step}
 d- h_m &= n +\delta h (\boldsymbol\theta_{\text{tr}}) + h_m(\boldsymbol\theta_{\text{tr}})-  h_m(\boldsymbol\theta_{\text{bf}})\nonumber\\
&\approx  n + \delta h (\boldsymbol\theta_{\text{bf}}) - \Delta\theta^i \partial_i  h_m (\boldsymbol\theta_{\text{bf}})\,,
\end{align}
where in the last line we take $\delta\vec h (\boldsymbol\theta_{\text{tr}})\approx \delta\vec h (\boldsymbol\theta_{\text{bf}})$. With all waveform models evaluated at the best-fit parameters, we deduce that~\citep{Cutler:2007mi}
\begin{align}
(\partial_i h_m|  d-h_m)
&\approx(\partial_i h_m|  n)+(\partial_i h_m| \delta h) - \Delta\theta^j\Gamma_{ij}=0\,,\nonumber\\
\Longleftrightarrow \Delta\theta^i &= (\Gamma^{-1})^{ij}\left[(\partial_j h_m| n)+(\partial_j h_m| \delta h)\right] \label{eq:CV_altorithm_multiple}\,.
\end{align}
where we now separate $\Delta\theta$ into an error from instrumental noise, $\Delta\theta_{\text{noise}}^i$ and a theoretical bias $\Delta\theta_{\text{sys}}^i$,
\begin{align}
\Delta\theta_{\text{noise}}^i &= (\Gamma^{-1})^{ij}(\partial_{j}h_{m}|n), \label{eq:width_noise}\\
\Delta\theta_{\text{sys}}^i& = (\Gamma^{-1})^{ij}(\partial_j h_{m}| \delta h). \label{eq:width_sys}
\end{align}
This expression for systematic errors first appeared in~\citep{Flanagan:1997kp}, see their Eq.~(6.11), although the implications were not studied in that paper. A much more through analysis was given in~\citep{Cutler:2007mi}. A geometrical intuition for the origin of Eqs.~(\ref{eq:width_noise},\ref{eq:width_sys}) is given in appendix~\ref{app:geometry}. A \texttt{python} tutorial on how to use equations \eqref{eq:width_noise} and \eqref{eq:width_sys} can be found \href{https://github.com/OllieBurke/Noisy_Neighbours/tree/main/Basic_CV_Formalism}{here}.

Generally speaking, a waveform model $h_{m}$ is ``good enough'' for parameter estimation if and only if $\Delta\theta^i_\text{sys}\lesssim \Delta\theta^i_\text{noise}$. The quantity $\Delta\theta^i_\text{noise}$ is a zero-mean random variable, so this inequality should hold in an average sense. %Inmeans that the theoretical bias should be less than the standard deviation of that random variable.
The $1\sigma$ deviation of $\Delta \theta_{\text{noise}}^{i}$ is $\Delta\theta^{i}_{\text{stat}} = \sqrt{\Gamma^{-1})^{ii}}$, so we define the function
\begin{equation}\label{eq:R_func}
    \mathcal{R}(\Delta\theta) := \rvert \Delta\theta^{i}/\Delta\theta_{\text{stat}}^{i}\rvert,
\end{equation}
and consider biases on the parameter $\theta$ arising from systematic effects to be significant %(namely exceed the width of the likelihood)
whenever $\mathcal{R}(\Delta\theta)>1$.
To conclude this section, we note that the statistical error $\Delta\boldsymbol\theta_\text{stat} \sim
\rho^{-1}$, while the systematic error $\Delta\boldsymbol\theta_\text{sys} \sim
\rho^{0}$. This implies that biases from modelling errors are independent of the SNR, while statistical errors become smaller as the SNR increases. Therefore, we expect systematics to become more important for loud sources.

\section{Generalisations} \label{sec:Source_Confusion_Bias}
%%%%%%%%%%%%%%%%%%%%%%%%%%%%%%%%%%%%%%%%%%%%%%%%%%

We now generalise the formalism represented by Eqs.~\eqref{eq:width_noise} and~\eqref{eq:width_sys} to two new cases, the first being the presence of \emph{confusion noise} from signals that have not been fitted for in parameter estimation, and the second being the inclusion of multiple signals in the data stream that are incorrectly modelled with approximate waveforms. 

\subsection{Source Confusion Bias}

The likelihood \eqref{eq:whittle_likelihood} only assumes that the noise $n(t)$ is both stationary and Gaussian (with zero mean). The noise $n(t)$ is usually assumed to be instrumental and modelled through the PSD via~\eqref{eq:Wiener-Khinchin-Theorem_freq}. However, in third-generation or space-based detectors there may be additional astrophysical contributions to the data stream from unresolved 
foregrounds of other GW signals~\citep{Crowder_2007,B_aut_2010,Robson:2017ayy,Roebber:2020hso,Korol:2020hay,Samajdar:2021egv,Pizzati:2021gzd,Karnesis:2021tsh}. 
This confusion noise $\Delta H_{\text{conf}}$ can be represented as part of the signal component of the data stream~\eqref{eq:data},
\begin{equation}\label{noisetot}
    d(t) = h_{e}(t;\boldsymbol{\theta}_{\text{tr}}) + n(t) + \Delta H_{\text{conf}}(t;\boldsymbol\theta^{(i)})\,.
\end{equation}
To understand when such confusion foregrounds can lead to biases, one may consider it to be 
a (deterministic) superposition of $N$ signals,
\begin{equation}\label{confres}
 \Delta  H_\text{conf} (t;\boldsymbol\theta^{(i)})= \sum_{i=1}^{N} h^{(i)}_e (t;\boldsymbol\theta^{(i)})\,.
\end{equation}
Equation~\eqref{eq:CV_1step} now becomes
\begin{equation}
    d- h_m = n + \Delta  H_\text{conf}+ \delta h (\boldsymbol\theta_{\text{tr}}) + h_m(\boldsymbol\theta_{\text{tr}})-  h_m(\boldsymbol\theta_{\text{bf}})\,,
\end{equation} 
from which we deduce the extra contribution to the biases~\eqref{eq:width_noise} and~\eqref{eq:width_sys} that originates from the source confusion term is
\begin{equation}\label{eq:bias_conf}
    \Delta\theta^{i}_{\text{conf}} = (\Gamma^{-1})^{ij}(\partial_j  h_m| \Delta H_\text{conf}).
\end{equation}
By analogy with~\eqref{eq:R_func}, source confusion from unfitted signals can be said to bias parameter estimates when its size exceeds the 1$\sigma$ deviations arising from instrumental noise fluctuations, which is true if $\mathcal{R}(\Delta\theta_\text{conf})>1$.
To summarise, when inferring the parameters of a single source, the total error is given by the sum of statistical error from noise fluctuations and the biases from source confusion and waveform errors through
\begin{equation}\label{eq:total_bias_one_source}
\Delta\theta^{i} = \Delta\theta_{\text{noise}}^{i} + \Delta\theta_{\text{sys}}^{i} + \Delta\theta_{\text{conf}}^{i},
\end{equation}
with the above terms from left to right given by Eqs.(\ref{eq:width_noise},\ref{eq:width_sys},\ref{eq:bias_conf}) respectively.  

In general, the confusion noise  contribution to \eqref{eq:total_bias_one_source}
depends on the particular sources from the unresolved population that are present in the data and so it is a random quantity. The correct way to handle this is to marginalise the likelihood of the corrected data stream, $d(t) - \Delta  H_\text{conf} (t;\boldsymbol\theta^{(i)})$, over the distribution of possible confusion backgrounds, $p(\Delta  H_\text{conf})$. This is a computationally expensive procedure and it is therefore difficult to obtain insights in that way. An alternative avenue to understanding when confusion is important, is to use the formalism described here to work with the bias induced by the confusion noise, $\Delta\theta_{\text{conf}}^{i}$, which is also a random quantity. %The confusion contribution can be treated as a statistic $\widecheck{\Delta\theta^{i}} = \Delta\theta^{i}_{\text{conf}}$, from which  one computes the expectation values related to its mean and variance, \andrea{changed $h_e$ into $\Delta H_{\text{conf}}$}
We can characterise it at the order of the linear signal approximation through its mean and variance. Since the total confusion noise contribution is a superposition of contributions from $N$ independent sources, the mean and variance of the total contribution is $N$ times the mean and variance of the contribution from a single source, $h_{e}(\boldsymbol{\theta}_{\text{conf}})$, which are
\begin{align}
    \mu_{\text{conf}}^i 
    %= \langle \Delta\theta_{\text{conf}}^{i}\rangle 
    %\Delta H_{\text{conf}}
    &= \int (\Gamma^{-1})^{ij}(\partial_j  h_m| h_e (\boldsymbol\theta_{\text{conf}})) \,p_{\text{pop}}(\boldsymbol\theta_{\text{conf}}) \, {\rm d} \boldsymbol\theta_{\text{conf}}\,, \label{eq:mu_conf}\\
    \Sigma_{\text{conf}}^{ij} &= \int (\Gamma^{-1})^{ik}(\partial_k  h_m| h_e (\boldsymbol\theta_{\text{conf}})) \times \nonumber \\
    & \hspace{1cm}(\Gamma^{-1})^{jl}(\partial_l  h_m| h_e (\boldsymbol\theta_{\text{conf}})) \,p_{\text{pop}}(\boldsymbol\theta_{\text{conf}}) \, {\rm d} \boldsymbol\theta_{\text{conf}} \nonumber \\
    & \hspace{3cm} - \mu_{\text{conf}}^i \mu_{\text{conf}}^j.\label{eq:sigma_conf}
\end{align}
Here, $p_{\text{pop}}(\boldsymbol\theta_{\text{conf}})$ is the probability density function of the population of confusion sources.
We would normally expect the mean to be close to zero, since for some sources in the population the bias would be positive and others negative and so it averages to zero (though this is not guaranteed to be the case). Regardless, the variance does not vanish, driving the total error to grow like a random walk as the total number of sources contributing to the confusion background increases. 

For large $N$, we can find a scaling relationship for the total bias using the central limit theorem 
\begin{align}
    &\frac{\sqrt{N}}{(N\Sigma_{\text{conf}})^{1/2}}(\Delta\theta_{\text{conf}} - N\mu_{\text{conf}}) \rightarrow \mathcal{N}(0,1) \\ &\Longrightarrow \Delta\theta^{i} - N \mu_{\text{conf}}^i \sim \sqrt{N} (\Sigma_{\text{conf}}^{ii})^{1/2}\,X,\label{eq:confnoiseth}
\end{align}
where $X$ is a standard Normal random variable. 
This behaviour will be investigated further in Sec.(\ref{sec:source_conf}).

In appendix \ref{App:confusion_noise}, we give a treatment of the confusion noise under the assumption $\Delta H_{\text{conf}}$ is a stationary time-series. In this prescription, making reference to the discussion above Eq.\eqref{eq:confvar}, the power of the confusion noise is folded into the PSD to form a combined noise PSD $S_{n}(f) \mapsto S_{n}(f) + S_{\text{conf}}(f)$. In realistic scenarios, due to the relative orientation of the galactic center with respect to the detector plane, the confusion noise will exhibit time-dependent amplitude modulations --- a non-stationary effect. In this work we will not treat $\Delta H_{\text{conf}}$ as a stationary time series and instead include it as an arbitrary superposition of sinusoids present in the data stream. We will treat both $n(t)$ and $\Delta H(t)$ as independent sources of noise and do not combine them into a single noise component $N(t)$.

\subsection{Biases due to waveform modelling errors}
We now generalise Equations~\eqref{noisetot} and \eqref{eq:total_bias_one_source} to the case of inference on multiple sources within the data stream. Similar ideas can be found in~\citep{Robson:2017ayy} for the case of massive black holes and galactic binaries in LISA. Here, we extend their discussion and include a prescription for the effect of waveform errors and confusion noise, generalising their results to multiple source types with an arbitrary number of sources.
We suppose there are $J$ different types of source in the data. We suppose that there are $N_j$ sources of type $j$ in the data stream, indexed by $i$, which each depend on a set of $m_j$ parameters, denoted by $\boldsymbol{\theta}^{(j)}_i$, which determine the corresponding gravitational waveform, $h^{j}(t;\boldsymbol{\theta}^{(j)}_i)$. The complete data stream can be written as
\begin{align}
d(t) &= h(t;\boldsymbol{\Theta}) + n(t) + \Delta H_{\text{conf}}\nonumber\\
&= \sum_{j=1}^J \sum_{i=1}^{N_j} h_{e}^{(j)}(t;\boldsymbol{\theta}^{(j)}_i) + n(t) + \Delta H_{\text{conf}}.
\label{eq:general_data}
\end{align}
Here we have introduced a composite vector of parameters, $\boldsymbol{\Theta} = \{\boldsymbol{\theta}^{(j)}_i\}^{j = 1,\dots,J}_{i = 1,\dots,N_{j}}$, such that $\Theta_{N_{<j}+(i-1)m_j+k} = (\boldsymbol{\theta}^{(j)}_i)_k$, where $N_{<j} = \sum_{l=1}^{j-1} N_l m_l$. 
For any given parameter in $\boldsymbol{\Theta}$, there is exactly one waveform in the above sum that depends on that parameter. Thus the derivatives of the signal reduce to derivatives of the specific waveform template. The combined Fisher matrix has a block structure, with the on-diagonal blocks being the Fisher matrices for the individual sources, and the off-diagonal blocks being formed from overlaps of waveform derivatives of one source with waveform derivatives of another source. Through calculating the Fisher matrix on parameters $\boldsymbol{\Theta}$, one is able to estimate the expected precision of  measurements on individual parameters, taking into account \emph{all} parameter correlations. This is (an estimate for) the precision that would be achieved in a simultaneous coherent fit to all sources in the data.

Without loss of generality, we illustrate this considering two classes of sources, with one source in the first class ($j=1$, $N_{1} = 1$) and an arbitrary number $N_{2}$ of sources in the second ($j=2$). This split is only made for ease of exposition, and is quite arbitrary as the sources could always be relabelled so that the first source is the source of interest. We want to estimate the impact of confusion due to the presence of the population of (fitted) sources of type 2, on the precision of parameter estimation for source 1. We define the following quantities
	\begin{align}
	&\Gamma^{(1)}_{jk}= \left(\partial_j  h^{(1)}(\boldsymbol\theta^{(1)})\big|\partial_k  h^{(1)}(\boldsymbol\theta^{(1)})\right)\label{eq:gamma_source_1_jk}\\
	& \left(\Gamma^{(2)}_i\right)_{jk} = \left(\partial_j  h^{(2)} (\boldsymbol\theta^{(2)}_i) \big| \partial_k h^{(2)} (\boldsymbol\theta^{(2)}_i)\right)\\
	&\left(\Gamma^{\rm mix}_i\right)_{jk}  = \left(\partial_j h^{(1)}(\boldsymbol\theta^{(1)})\big|\partial_k  h^{(2)}(\boldsymbol\theta^{(2)}_i)\right)\label{eq:gamma_source_mix}.
	\end{align}
Here $\Gamma^{(1)}$ is the Fisher matrix for the source of type 1, $\Gamma^{(2)}_i$ is the Fisher matrix for the $i$'th source of type 2 ($i = 1,\dots, N_2$) and $\Gamma^{\rm mix}_i$ is the mixed Fisher matrix for the source of type 1 and the $i$'th source of type 2.
In what follows, we find it useful to combine the Fisher matrix contributions of the entire population of sources in a more compact form. One can write Eqs.(\ref{eq:gamma_source_1_jk}-\ref{eq:gamma_source_mix}) as
\begin{align}
    &\left(\Gamma^{(2)}\right)_{m_2(i-1)+j,m_2(l-1)+k}=\left(\partial_j  h^{(2)}(\boldsymbol{\theta}^{(2)}_{i}) \big| \partial_k  h^{(2)}(\boldsymbol{\theta}^{(2)}_{l})\right) \label{eq:convoluted_fish_matrix}\\
	&\Gamma^{\rm mix}_{j,m_2(i-1)+k} =(\partial_j  h^{(1)}(\boldsymbol\theta^{(1)})|\partial_k  h^{(2)}(\boldsymbol{\theta}^{(2)}_{i})).\label{gammamix}
\end{align}
The Fisher matrix for the full analysis and its inverse are therefore
\begin{equation}\label{eq:inv_joint_matrix}
\Gamma = \left( \begin{array}{cc} \Gamma^{(1)}&\Gamma^{\rm mix} \\ (\Gamma^{\rm mix})^T&\Gamma^{(2)}\end{array}\right); \quad 
\Gamma^{-1} = \left( \begin{array}{cc} \Gamma^{-1}_{11}&\Gamma^{-1}_{12} \\ (\Gamma^{-1}_{12})^T&\Gamma^{-1}_{22}\end{array}\right)
\end{equation}
with the components of the inverse\footnote{We note also that
 	\begin{align*}
 	\Gamma^{-1}_{11} &= (\Gamma^{(1)})^{-1} +  (\Gamma^{(1)})^{-1}  \Gamma^{\rm mix}\Gamma^{-1}_{22}\,  (\Gamma^{\rm mix})^{T} (\Gamma^{(1)})^{-1} \\
 	\Gamma^{-1}_{22} &= (\Gamma^{(2)})^{-1} +  (\Gamma^{(2)})^{-1}  \Gamma^{\rm mix}\Gamma^{-1}_{11}\,  (\Gamma^{\rm mix})^{T} (\Gamma^{(2)})^{-1} 
 	\end{align*}
which can sometimes be cheaper to compute than Eq.~\eqref{eq:fish_matrix_22}.}
\begin{align}
    \Gamma^{-1}_{11} &= \left( \Gamma^{(1)} - \Gamma^{\rm mix} (\Gamma^{(2)})^{-1} (\Gamma^{\rm mix})^T\right)^{-1}\,, \label{eq:fish_matrix_11}\\
    \Gamma^{-1}_{22} &= \left( \Gamma^{(2)} - (\Gamma^{\rm mix})^T (\Gamma^{(1)})^{-1} \Gamma^{\rm mix}\right)^{-1}\,, \label{eq:fish_matrix_22}\\
    \Gamma^{-1}_{12} &= -\Gamma^{-1}_{11}\, \Gamma^{\rm mix} (\Gamma^{(2)})^{-1}\,. \label{eq:fish_matrix_12}
\end{align}
The components $\Gamma^{-1}_{11} $ encode the measurement precisions for source 1. If the degree of correlation between the source types is small, i.e., $|\Gamma^{\rm mix}| \ll 1$, we can approximate this as
\begin{equation}
\Gamma^{-1}_{11} \approx (\Gamma^{(1)})^{-1} + (\Gamma^{(1)})^{-1} \Gamma^{\rm mix} (\Gamma^{(2)})^{-1} (\Gamma^{\rm mix})^T (\Gamma^{(1)})^{-1}. \label{eq:coverrorapprox}
\end{equation}
The first term is the measurement precision when there are no sources in the data, while the second represents the degradation in the precision due to confusion with the other sources. We can understand the form of the second term as follows. If the other sources were ignored when fitting for source 1, the parameter bias would be given by Eq.~\eqref{eq:width_sys}
\begin{equation}
\Delta \theta^{(1),i}_{\rm sys} = (\Gamma^{(1)})^{-1}_{ij} (\partial_j h_m^{(1)} | \boldsymbol{h}^{(2)})
\label{eq:explainbias}
\end{equation}
where we are combining all of the sources of type 2 into the single term $\boldsymbol{h}^{(2)}$. This bias is dominated by the contribution from the true waveform. When we simultaneously fit for the sources of type 2, we imperfectly remove these signals, leaving a residual in the data of the form $\partial_j \boldsymbol{h}^{(2)} \Delta \theta_2^j$, where again we are combining the parameters of all of the sources of type 2 into a single parameter vector, $\boldsymbol{\theta}_2$. The parameter error, $\Delta \boldsymbol{\theta}_2$, is a random variable with covariance matrix $\langle \Delta \theta_2^j \Delta \theta_2^k \rangle = (\Gamma^{(2)})^{-1}_{jk}$. The bias on source 1 parameters can be approximated by $\Delta \theta^{(1),i}_{\rm sys} \approx (\Gamma^{(1)})^{-1}_{ik}(\partial_{k}h^{(1)}|\partial_{l}\boldsymbol{h}^{(2)}\Delta\theta_{2}^{l})$. The covariance of the induced systematic error in the parameters of source 1 is then 
\begin{align*}
    \langle \Delta \theta^{(1),i}_{\rm sys} \Delta \theta^{(1),j}_{\rm sys} \rangle &= (\Gamma^{(1)})^{-1}_{ik} (\partial_k h_m^{(1)} |\partial_l \boldsymbol{h}^{(2)}) \langle \Delta \theta_2^l \Delta \theta_2^m \rangle \\
    & \hspace{1cm} (\partial_n h_m^{(1)} |\partial_m \boldsymbol{h}^{(2)}) (\Gamma^{(1)})^{-1}_{jm} \\
    &=\left[(\Gamma^{(1)})^{-1} \Gamma^{\rm mix} (\Gamma^{(2)})^{-1} (\Gamma^{\rm mix})^T (\Gamma^{(1)})^{-1}\right]_{ij},
\end{align*}
which is the second term from Eq.~\eqref{eq:coverrorapprox}. 
There is nothing that can be done to mitigate uncertainties of this type, which arise from an over-abundance of sources in the data. However, as described above, additional uncertainties can arise from their inaccurate modelling.
%of the gravitational wave signals. 
Previous studies have focused on biases from inaccurate modelling of the target source, but it is also important to ask if the inaccurate modelling of a large number of other sources can leave a sufficient residual in the data to cause problems.

To estimate this, we define $\delta h^{(1)} =  h^{(1)}_e -  h^{(1)}_m$ as the difference between the exact $h_e$ and template $h_m$ waveforms for the source of type 1, and similarly $\delta h^{(2)}_i = h^{(2)}_e(\boldsymbol\theta^{(2)}_i) -  h^{(2)}_m(\boldsymbol\theta^{(2)}_i)$ for the $i$'th source of type 2. We also define $\delta h = \delta h^{(1)} + \sum_{i=1}^{N_2} \delta h^{(2)}_i$ as the combination of all waveform residuals. %These template errors affect parameter biases equally. 
Let us define the bias vector $\boldsymbol{b}$ 
\begin{multline*}
    \boldsymbol{b} = (b^{(1)}_{1},\ldots,b^{(1)}_{m_{1}},(b^{(2)}_{1})_{1},\ldots, \\ (b_{1}^{(2)})_{m_{2}},\ldots,(b^{(2)}_{N_{2}})_{1}, \ldots, (b^{(2)}_{N_{2}})_{m_{2}})^{T},
\end{multline*}
such that $\boldsymbol{b} = [\boldsymbol{b}^{(1)},\boldsymbol{b}^{(2)}]\in\mathbb{R}^{(m_{1} + N_{2}m_{2})\times 1}$ with individual components given by
\begin{align}
b^{(1)}_j &= (\partial_j  h^{(1)} (\boldsymbol\theta^{(1)})| \delta  h ),  \\
(b^{(2)}_i)_j &= (\partial_j h^{(2)}(\boldsymbol\theta^{(2)}_i) | \delta  h ). \nonumber 
\end{align}
Note that the bias defined here is only the contribution from modelling errors. The full shift in the peak of the likelihood may be found from a similar expression, with $n(t)$ and $\Delta H_{\text{conf}}$ added to $\delta h$ in the inner products. The quantity $b^{(1)}_{j}$ for $j = 1,\ldots,m_{1}$ are the components $\boldsymbol{b}$ for the first source of type 1. The quantity $(b_{i}^{(2)})_{j}$ are the $j$th components of $\boldsymbol{b}$ with respect to the $i$th source of type 2. The vector $\boldsymbol{b}$ can be written more concisely as
\begin{align}
& b_j =  b^{(1)}_j \quad \mbox{for } j=1,\ldots,m_1, \label{eq:b_vec_sys_1}\\ 
& b_{m_1+m_2(i-1)+j} = (b^{(2)}_i)_j \mbox{ for } i=1,\ldots,N_2; j=1,\ldots,m_2,\label{eq:b_vec_sys_2}
\end{align}
The biases computed from Eq.~\eqref{eq:width_sys} are given by $\Delta\boldsymbol\Theta = \Gamma^{-1}\boldsymbol{b}$ and are thus
\begin{equation}\label{eq:master}
\Delta \boldsymbol\Theta := \begin{pmatrix}
\Delta\boldsymbol\theta^{(1)}  \\[6pt]\Delta\boldsymbol\theta^{(2)}
\end{pmatrix}=\Gamma^{-1} \begin{pmatrix}\boldsymbol{b}^{(1)}  \\[6pt]\boldsymbol{b}^{(2)}
\end{pmatrix} 
\,,
\end{equation}
Using Eqs.(\ref{eq:convoluted_fish_matrix}-\ref{eq:inv_joint_matrix}) and Eqs.(\ref{eq:b_vec_sys_1},\ref{eq:b_vec_sys_2}), the bias in the source parameters of the signal of type 1 is
\begin{equation}\label{eq:source_1_bias}
\Delta \theta^{(1)}_i = (\Gamma^{-1}_{11})^{ij}  b_{j} + (\Gamma^{-1}_{12})^{im} b_{m_1+m}\,,
\end{equation}
with components of $(\Gamma^{-1}_{11})$ and $(\Gamma^{-1}_{12})$ defined in Eqs.(\ref{eq:fish_matrix_11},\ref{eq:fish_matrix_12}). Using the approximation that led to Eq.~\eqref{eq:coverrorapprox}, that the elements of $\Gamma^{\rm mix} (\Gamma^{(2)})^{-1} (\Gamma^{\rm mix})^T$ are much smaller than those of $\Gamma^{(1)}$, we can approximate Eq.~\eqref{eq:source_1_bias} as
\begin{multline}\label{multibias}
\Delta \theta^{(1)}_i \approx [(\Gamma^{(1)})^{-1}]^{ij} b_{j} - \\ [(\Gamma^{(1)})^{-1}]^{ij} (\Gamma^{\text{mix}})_{jl} [(\Gamma^{(2)})^{-1}]^{lm} b_{m_1+m} 
%  + \mathcal{O}((\Gamma^{\text{mix}})^{ij} (\Gamma^{(2)})_{jl}^{-1} (\Gamma^{\text{mix}})_{lp}^T.
\end{multline}
We see that there are two contributions to the parameter bias on the single source of type 1: the standard CV bias \eqref{eq:width_sys} arising from mismodelling of that source; and an extra correction due to mismodelling of overlapping sources. If the sources from each source type are orthogonal, $\Gamma^{\text{mix}}\rightarrow 0$, then the presence of other sources does not contribute a parameter bias.

In testing the formalism below, we drop the source type indices for simplicity. 
The waveform and shift in the peak of the likelihood will be denoted%are given by
\begin{align}
  h(t;\boldsymbol\Theta) &=\sum_{i=1}^{N} h_{e}(t;\boldsymbol{\theta}_i)\quad \text{for $\boldsymbol{\Theta} = \{\boldsymbol{\theta}_{1},\ldots,\boldsymbol{\theta}_{N}\}$}\nonumber\,,\\
  \Delta\Theta^{i}&=(\Gamma^{-1})^{ij}\left(\frac{\partial h}{\partial \Theta^{j}}\bigg| n(t) + \delta h + \Delta H_{\text{conf}} \right)\nonumber \\
  &= \Delta\Theta^{i}_\text{noise} + \Delta\Theta^{i}_\text{sys} + \Delta\Theta^{i}_\text{conf}\label{multipar}\,,
\end{align}
with total theoretical error $\delta h$ and Fisher matrix denoted
\begin{equation*}
    \Gamma_{ij} = \left(\frac{\partial h}{\partial \Theta^{i}}\bigg\rvert \frac{\partial h}{\partial \Theta^{j}}\right) \label{eq:FM_one_source_type},\quad  \delta h= \sum_{i = 1}^{N}(h_{e}^{(1)}(t;\boldsymbol{\theta}_{i}^{(1)}) - h_{m}^{(1)}(t;\boldsymbol{\theta}_{i}^{(1)})).
\end{equation*}
The $\Gamma$ appearing in~\eqref{multipar} is the joint Fisher matrix $\Gamma \in \mathbb{R}^{(N\times m)\times(N\times m)}$, with $m$ the dimension of each parameter space $\boldsymbol{\theta}_{1},\ldots, \boldsymbol{\theta}_{N}$.
Equation~\eqref{multipar} is separated into a noise induced error, $\Delta\Theta^{i}_\text{noise}$, and biases split into a confusion noise contribution, $\Delta\Theta^{i}_\text{conf}$, and a contribution from theoretical errors, $\Delta\Theta^{i}_\text{sys}$. From Eq.\eqref{eq:R_func}, biases are then significant whenever $\mathcal{R}(\Delta\Theta^{i}_\text{conf} + \Delta\Theta^{i}_\text{sys})>1$.

\section{Modelling signals and noise}\label{sec:model&noise}

To illustrate the above formalism, we will consider a number of simplified scenarios. For all of these we will model the signals using the \texttt{TaylorF2} waveform model
\begin{align}\label{eq:signal_model_SPA}
    \hat h(f) &= \mathcal{A}\left(\frac{\pi G M f}{c^3}\right)^{-7/6}e^{-i\psi(f)}\,,  \\
    \mathcal{A} &= -\sqrt{\frac{5}{24}}\frac{c}{D_\text{eff}\pi^{2/3}}\left(\frac{G\mathcal{M}_c}{c^3}\right)^{5/6}\,.
\end{align}
Here, $M_c:=M \eta^{3/5}$ is the chirp mass and $D_\text{eff}$ the effective distance.
For this reason $D_\text{eff}$ should be treated effectively as an overall scaling factor, and not as a physical distance parameter. 
We retain only the leading-order amplitude $\mathcal{A}$~\citep{Allen:2005fk} in the waveform. The phase is PN-expanded in the velocity $v:=(\pi M G f/c^3)^{1/3}$ and reads
\begin{equation}
    \psi(f) = 2 \pi  f t_c - \phi_c+\frac{3v^{-5}}{128 \eta} \left(1 +\sum_{n=2}^{n=7} v^n \psi_{\frac{n}{2}\text{PN}}\right)\,,
\end{equation}
with coefficients up to 3.5PN as given in Sec.IIIB of \citep{Cutler:2007mi}.
The constant portion of the phase depends on the time and phase at coalescence, $t_c$ and $\phi_c$. We have only included spin-orbit interactions in the 1.5PN phase through the spin parameter $\beta$, defined in~\citep{Berti:2004bd}. We remark that $\beta$ satisfies the inequality $|\beta|\lesssim 9.4$.
We take the above \texttt{TaylorF2} model to be the exact waveform $\hat h_e(f;\boldsymbol\theta)$.
In these examples, for simplicity we will treat the phase, $\phi_c$, time of coalescence, $t_{c}$, and distance, $D_\text{eff}$, as perfectly-known parameters. 
Notice that we also ignore the effect of the detector response function.
Ignoring the detector response is a restrictive simplification, since over the observation time in either ground-based or spaceborne detectors we would expect the phase and amplitudes of the signal to be modulated by detector motion. Moreover, the angular dependence introduced by the detector response leads to a multi-modal and generally non-gaussian likelihood \citep{Cornish:2020vtw,Marsat:2020rtl}, which our Fisher matrix cannot reproduce. 
As
the purpose of our examples is to illustrate the formalism of Sec.(\ref{sec:Source_Confusion_Bias}) the simplifications we make here are not a serious restriction, though the impact must be assessed in future studies.

To evaluate the modelling error we need an estimate for the waveform uncertainty, which is necessarily not known exactly. If this is completely unconstrained, then modelling errors lead to non-estimable ``stealth biases'' in waveform parameters \citep{Vallisneri:2013rc}. However, in practice we generally have an idea of how large modelling uncertainties are, by comparing two different waveform models, or two different orders of expansion of the same waveform model. Given an estimated waveform difference, we can use the formalism described here to assess if that model is good enough to avoid significant systematic errors in parameter estimation.
To represent modelling inaccuracies, we represent the approximate waveform by modifying the smallest contribution in the 3.5PN phase contribution
\begin{equation}
    \psi_{\text{3.5PN}}^{(\epsilon)}:=\pi  \left(\frac{77096675}{254016}+\frac{378515}{1512}\eta -(1-\epsilon)\frac{74045}{756}\eta ^2\right)\,,
\end{equation}
(for $\epsilon \in [0,1]$). The true PN  waveform  has $\epsilon=0$  and we will take a (fixed) value of $\epsilon \neq 0$ to represent the approximate model. 
Finally, we model confusion noise as a superposition of \texttt{TaylorF2} models, unless otherwise specified (see Sec.~\ref{sec:results}).

We generate detector noise in both ET and LISA using Eq. \eqref{eq:Wiener-Khinchin-Theorem_freq} and the PSDs found in~\citep{Robson_2019} (LISA) and~\citep{regimbau2012mock} (ET). 
More details on how we generate our signals and noise realisations are found in Appendix~\ref{app:num_routines}. In Appendix~\ref{app:Fisher_Matrix} we describe how the waveform derivatives (and Fisher matrices) are calculated, and outline the MCMC techniques used to verify them. 

%%%%%%%%%%%%%%%%%%%%%%%%%%%%%%%%%%%%%%%%%%%%%%%%%%
\section{Results} \label{sec:illustrations}
%%%%%%%%%%%%%%%%%%%%%%%%%%%%%%%%%%%%%%%%%%%%%%%%%%

%With the formalism now in place, we are in a position to start exploring its features and applicability. 
In this section, we present four illustrations for the formalism described in Sec.~\ref{sec:Source_Confusion_Bias}. The first one concerns confusion and detector noise only. The second concerns the overlap of two signals with coincident coalescence. The third concerns theoretical errors from incorrectly removed waveforms only. The fourth considers all of the above combined.

\subsection{Biases from detector and confusion noise}\label{sec:source_conf}

\begin{figure*}
    \centering
    \includegraphics[height = 7.5cm, width = 17cm]{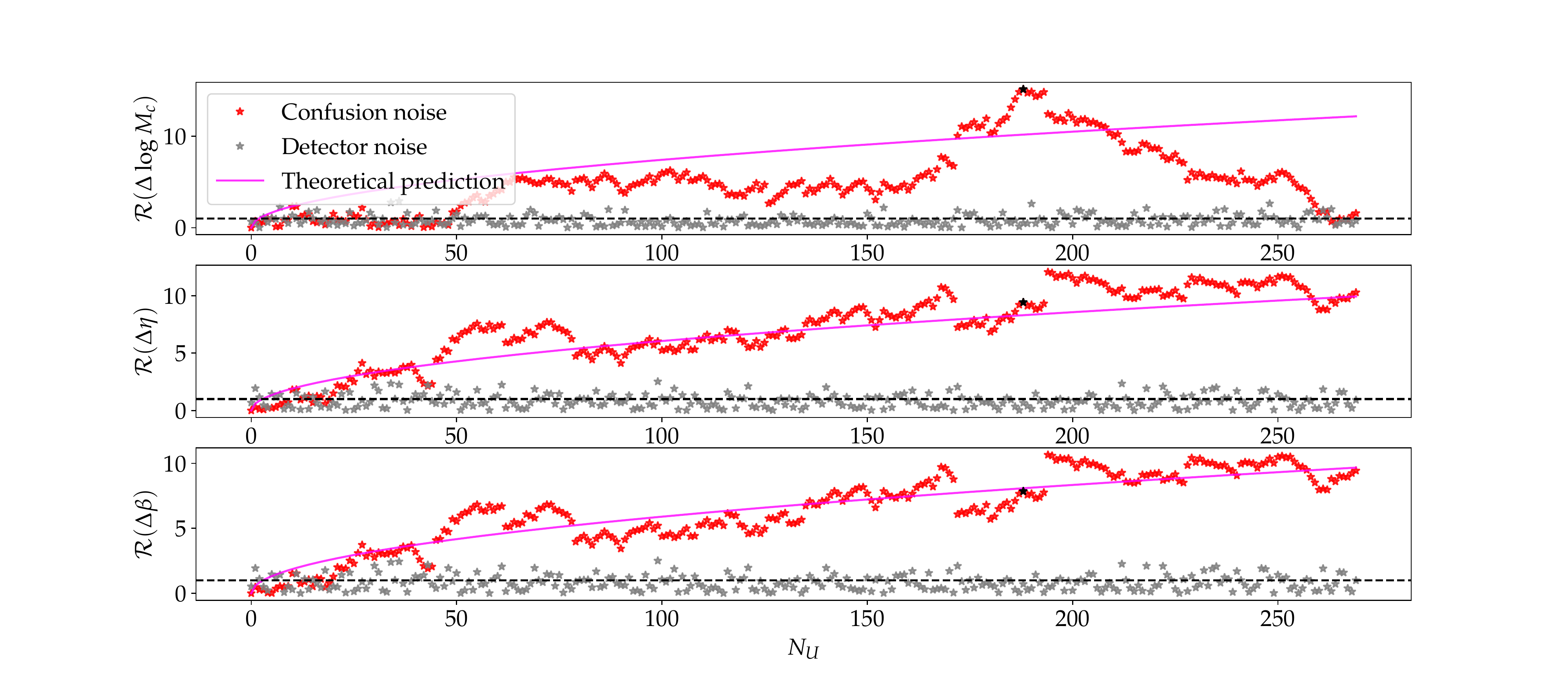}
    \caption{\emph{Accumulation of bias from population of overlapping signals}. In red, the accumulation of bias on the parameters of the reference signal from massive black hole binaries that have not been resolved. In gray, the statistical errors arising from instrumental noise fluctuations. The noise is independently generated for each data set and so we expect the ${\cal R}$ values to follow a $N(0,1)$ distribution, which is consistent with what is seen in the figure. In purple, the theoretical prediction, which follows a $\sqrt{N}$ behaviour according to Eq.\eqref{eq:confnoiseth}. In black, the data point with the largest bias in ${\cal M}_c$, for which the results were verified using an MCMC simulation, giving the posterior shown in Figure~\ref{fig:conf_noise_bias}. We note that these panels are not independent, as they represent one-dimensional marginals of a three dimensional distribution that has large correlations. 
    }
    \label{fig:conf_noise_LISA}
\end{figure*}

In this exploration, we consider a single reference signal in the LISA band and a confusion noise $\Delta H_\text{conf}$ of binaries that follow a realistic mass distribution. Our aim is to understand how much the combined effects of the confusion signals affect recovery of the parameters of the reference signal, and whether we can predict the biases using the formalism described above. The data stream we consider is
\begin{equation}\label{sig}
    \hat d(f) = \hat h_e(f;\boldsymbol\theta_\text{tr})+ \Delta H_\text{conf}(f;\boldsymbol\theta)+ \hat n(f).
\end{equation}
We recover the reference signal perfectly by modelling it with the exact waveform $\hat h_e(f; \boldsymbol\theta_\text{tr})$ of~\eqref{eq:signal_model_SPA} in both the Fisher matrix and the MCMC sampling algorithms. We therefore expect no biases from modelling errors.
We use the following configuration of true (injected) parameters, 
\begin{equation}
    \boldsymbol{\theta}_\text{tr} = \{\log \mathcal{M}_c = 83.34, \eta = 0.210, \beta = 5.00\}\,,
\end{equation}
which correspond to a spinning binary of total mass $M = 2\times 10^6 M_\odot$. We complete the full set of parameters by choosing an effective distance $D_\text{eff} = 1$Gpc and phase at coalescence $\phi_c = 0$, with time at coalescence given by the chirp time (see appendix~\ref{app:num_routines}). We begin observing the binary at $f_0=0.25\, m$Hz and stop at $f_\text{max} = 2.2\, m$Hz, corresponding to the ISCO frequency in a Schwarzschild spacetime for the chosen total mass. That is, we observe the binary until it chirps $\sim$4.4 days after we have started observing it. These choices lead to an SNR of $\rho\sim 4200$ for this signal, for which we expect the Fisher formalism to be a very good approximation.

To construct $\Delta H_\text{conf}(f; \boldsymbol\theta)$, we first build a mock catalogue of $N=800$ sources, which are sampled from uniform distributions
\begin{align}
    %D_\text{eff}&\sim U[1,5]\text{ Gpc}\nonumber\\
    \beta&\sim U[0.001,9.4] \nonumber\\
    \eta&\sim U[0.001,0.25] \nonumber\\
    \phi_c&\sim U[0, 2\pi]
\end{align}
and $t_c$ given by the individual chirp times. We distribute sources uniformly in volume by sampling distances $D_\text{eff}^3\sim U[1,125]\text{ Gpc}^3$.
We let the total masses of the binaries in this catalog follow a standard probability density function for massive black holes~\citep{Gair:2010bx,Gair:2010yu,Sesana:2010wy},
\begin{equation}
    \frac{dN}{dM} = \frac{\alpha\, M^{\alpha - 1}}{M^{\alpha}_\text{max}-M^{\alpha}_{\text{min}}},
\end{equation}
where the masses' range is $M_\text{min}=10^4M_\odot < M< 10^7 M_\odot=M_\text{max}$ and $\alpha=0.03$ is the fit in Ref.~\citep{Gair:2010yu} to the inactive massive black holes of~\citep{Greene:2007xw}. 
We can directly sample the total masses using
\begin{equation}
    \log M =\alpha^{-1} \log\big[(M^{\alpha}_\text{max}-M^{\alpha}_{\text{min}})\, u +M^{\alpha}_{\text{min}}\big],
\end{equation}
with $u \sim U[0,1]$. 
For each element of the catalogue, we compute the waveform of the binary using the exact model $h_e$. For those mass draws for which the frequency array of the binary is longer than that of the reference signal, we cut the former to be of the same length as the latter. Otherwise, we stop the evolution of the binary at its ISCO to avoid introducing an artificial portion of the waveform into the analysis. If the waveform has an observed SNR $\rho_\text{obs} = \rho + N(0,1)$~\citep{Sathyaprakash:2009xs}, where $N(0,1)$ is a standard normal distribution, such that
$\rho_\text{obs}  < \rho_\text{threshold} = 15$
then we consider the binary as ``missed'', retain the waveform and add it to $\Delta H_\text{conf}$ in a cumulative fashion. In our example, for $N=800$ events in the mock catalogue, $N_\text{U}=\mathcal{O}(270)$ have SNRs below the threshold and are thus unresolved. The final SNR of $\Delta H_\text{conf}$ is $\sim$170 in this case. 

Once $\Delta H_\text{conf}$ is obtained and the data stream~\eqref{sig} is thus fully specified, we predict the biases from confusion noise $\Delta\boldsymbol\theta_{\text{conf}}$ [namely, using~\eqref{multipar} retaining only $\Delta H_\text{conf}$ in the bias vector], which we can compare to the statistical error $\Delta\boldsymbol\theta_{\text{noise}}$ [found from~\eqref{multipar} with $n$ only]. We show the accumulation of the biases from confusion noise in Fig.~\ref{fig:conf_noise_LISA} by plotting the ratio $\mathcal{R}(\Delta\boldsymbol\theta_{\text{conf}})$. In this plot, calculations with different numbers of sources use different noise realisations, but consistent source catalogues, i.e., the data set with $N+1$ confusion sources includes the same sources as the $N$ confusion sources data set, plus one additional source. The ratio, $\mathcal{R}(\Delta\boldsymbol\theta_{\text{noise}})$, of the noise-induced shift in the peak of the likelihood to the expected standard deviation of this quantity, is  also  shown  and can be seen to hover around the value of $1$, as expected. Conversely, we find that the formalism predicts significant biases ($\mathcal{R}>1$) from the accumulation of missed signals drawn from a simple, but realistic distribution of the masses. We plot the theoretical prediction from Eq.~\eqref{eq:confnoiseth} of Sec.~\ref{sec:Source_Confusion_Bias} on top of the found ratios, showing that they (qualitatively) follow the expected $\sqrt{N}$ behaviour. We note that we do not expect the bias to precisely track the theoretical prediction. As sources are added the bias follows a random walk, and Eq.~\eqref{eq:confnoiseth} gives an approximate 1-$\sigma$ boundary to that random walk. We have tried many confusion noise realisations with $N_{U} \gg 1000$ and in all cases the accumulation of the bias follows a similar pattern. The realisation used in this figure happens to track the theoretical prediction quite well, but is reasonably typical.

\begin{figure}
	\includegraphics[width=\linewidth]{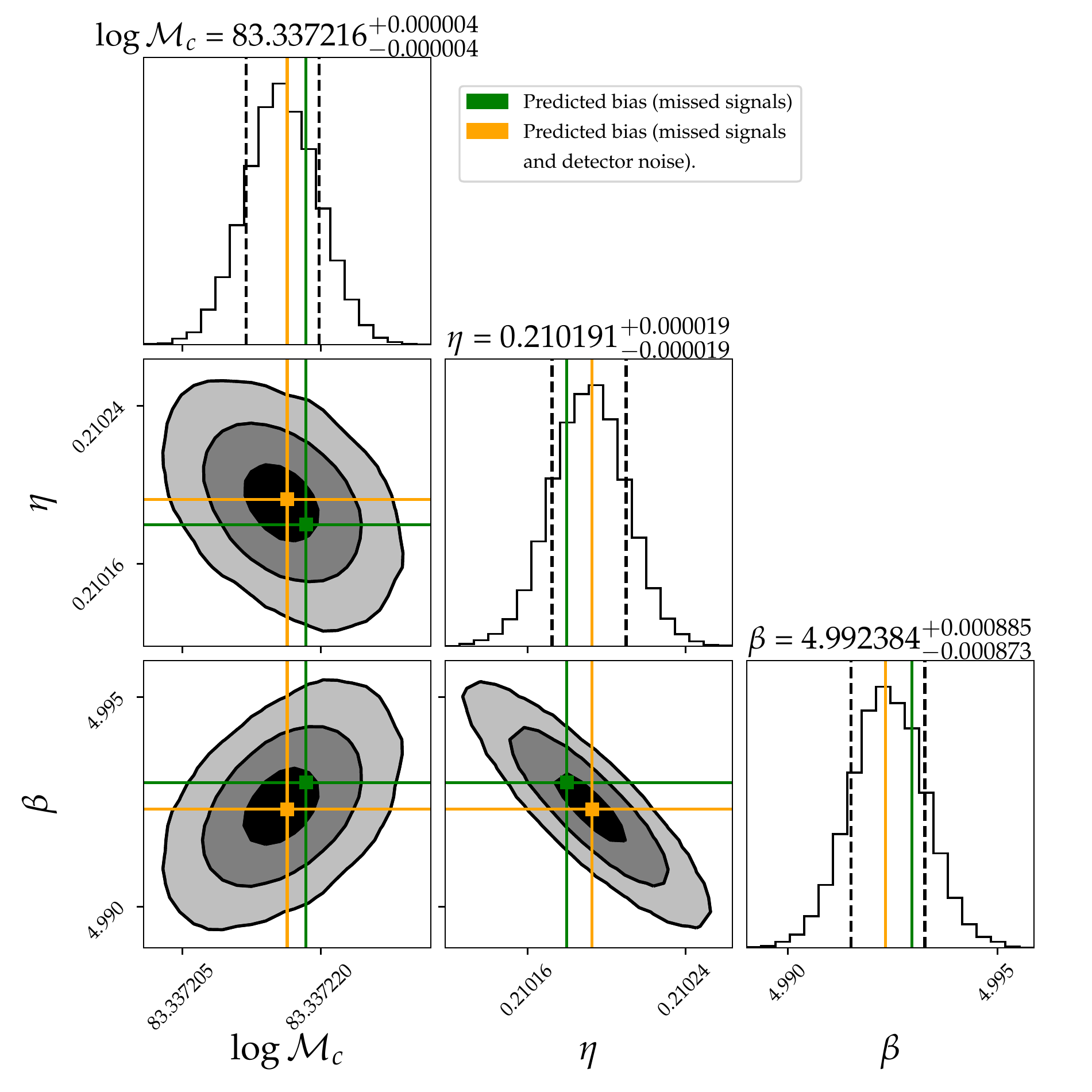}
	\caption{\emph{Biases from source confusion and detector noise}. MCMC posteriors and predictions from the Fisher formalism for the largest-bias case in Fig.~\ref{fig:conf_noise_LISA}. The values in green are predictions considering source confusion only. Those in orange combine biases from source confusion and detector noise (which we cannot access in a realistic situation). The true values are well beyond the range of the plot at $\sim 15 \sigma$ for each parameter (see Fig.~\ref{fig:conf_noise_LISA}). }
	\label{fig:conf_noise_bias}
\end{figure}

To assess whether these predictions are sound, we confirm them with an MCMC analysis for the data set that gives the largest bias (${\cal R} \sim 15$) in chirp mass, indicated by the black data point in Fig.~\ref{fig:conf_noise_LISA}). The result of the MCMC run and the predictions for the shift in the peak of the likelihood due to the confusion sources and noise, computed with  Eq.~\eqref{multipar}, are shown in Fig~\ref{fig:conf_noise_bias}. Even in this most extreme case, we can clearly see that the predictions for the bias match the MCMC posterior very well, demonstrating that the formalism works well in estimating source confusion from missed signals.
We remark that in this example the SNR of the residuals is lower than the SNR of the signal we are inferring from the data stream. This is a regime in which we would expect that the linear signal approximation is valid. In scenarios in which the SNR of the ``missed'' signals is larger than that of the target source, the linear signal approximation might cease to be valid, but this formalism should at least provide an indication that systematic biases are ``large''.

\subsection{Biases from overlapping signals with coincident coalescence}
\label{subsec:overlap_merger}

\begin{figure*}
	\includegraphics[width=\linewidth]{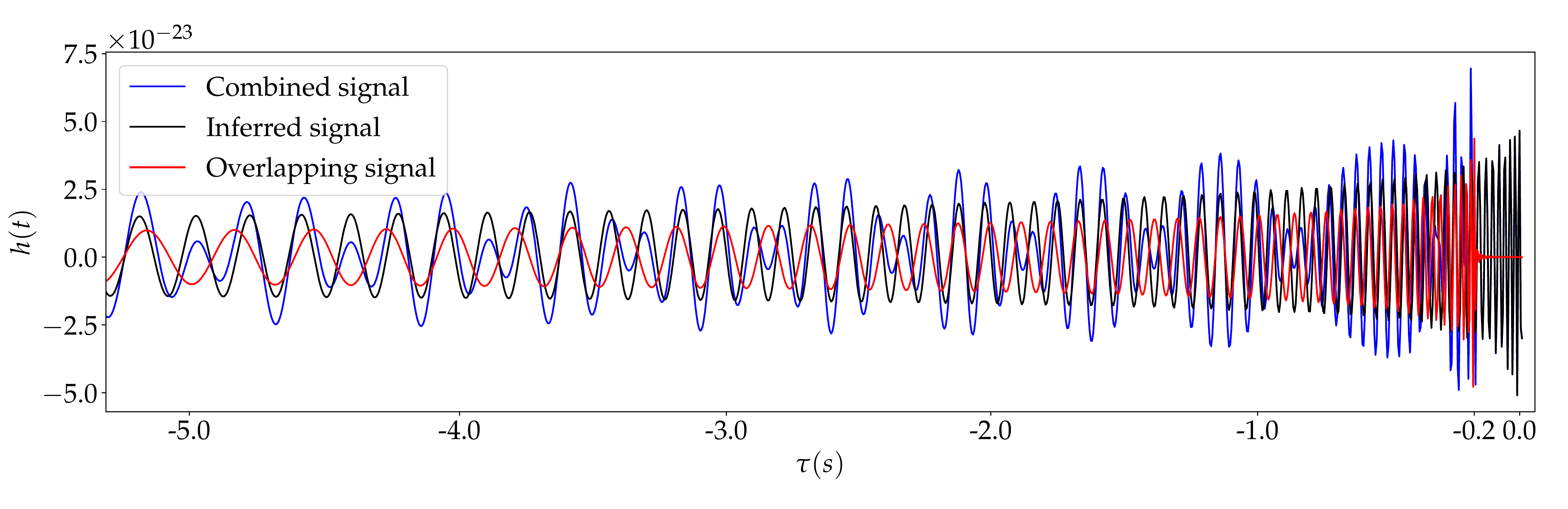}
	\caption{\emph{Waveforms for overlapping signals}. We plot the waveforms for signal $h_e^{(1)}(t)$ in black; this represents the ``inferred source'' for which we are attempting to recover the parameters. We plot the waveform of the overlapping signal $h_e^{(2)}(t)$ in red; the signal has a coalescence time at $\tau = -0.2s$ relative to the one of the inferred source. The sum of the two signals is shown in blue.}
	\label{fig:TD_waveforms}
\end{figure*}
A particularly interesting class of overlapping signals that has attracted attention in the recent literature are those where the coalescence times $t_c$ are nearly simultaneous. Such a scenario could be relevant to mergers of massive black holes observed by LISA or to stellar-origin binary black-holes (BBH) observed by ET and Cosmic Explorer (CE), but this will depend on the rate of such mergers and, therefore, the probability that mergers happen within the same time period. Quantitative studies of the rate of overlapping mergers have been carried out for advanced LIGO and CE. In~\citep{Samajdar:2021egv,Pizzati:2021gzd,Himemoto:2021ukb,Relton:2021cax}, the authors conclude that coincident (meaning merger times within 2 seconds) mergers of BBH binaries will occur tens of times per year for CE, and binary neutron star (BNS) mergers could occur coincidentally with other BNS or BBH mergers hundreds or even thousands of times per year.

The same papers, as well as~\citep{1853026}, also present the first Bayesian inference analyses with overlapping signals, with some critical differences. \citep{1853026} studies the simultaneous inference of overlapping neutron star binaries, in such a way that no biases on the parameters are expected from confusion noise. \citep{Relton:2021cax} performs a similar analysis for the second-generation LIGO-Voyager detector, \citep{Samajdar:2021egv} for pairs of BBH-BBH, BBH-BNS and BNS-BNS systems using  \texttt{LAL-inference}~\citep{Veitch:2014wba}, and \citep{Pizzati:2021gzd} for BBH pairs with \texttt{bilby}~\citep{Ashton:2018jfp}. However, in these last two papers, inference is performed for one binary only, treating the second as confusion noise. They find that biases occur when the difference between the coalescence times $\tau=t_c^{(2)}-t_c^{(1)}$ of signals ``(1)'' and ``(2)'' is sufficiently small, roughly $\tau\lesssim 0.5s$. Here we analyse a similar scenario to that of \citep{Pizzati:2021gzd}, interpreting the bias as arising from a single confusion source, to see whether the analytic formalism presented here can reproduce that result without the need for expensive Bayesian posterior computation. Notice that a (joint) Fisher-matrix analysis is presented in \citep{Himemoto:2021ukb} for a similar scenario, though the similarities end there.

\begin{figure}
	\includegraphics[width=\linewidth]{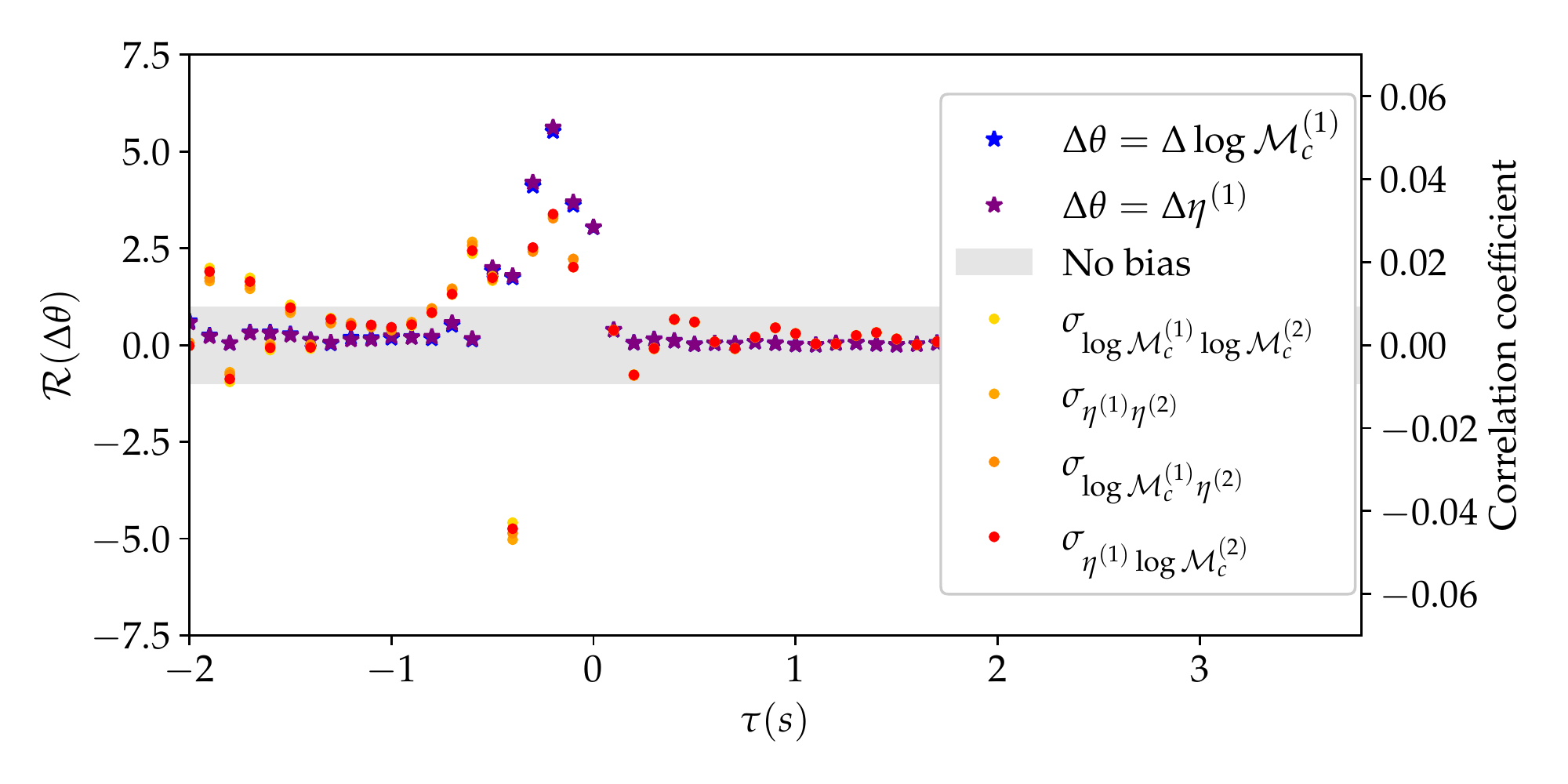}
	\caption{\emph{Biases from an overlapping signal as a function of the difference in coalescence time}. In cold (blue, purple) colors, we plot the bias ratios for the parameters of signal ``(1)'' due to the unaccounted-for presence of signal ``(2)'', as a function of the coalescence time difference $\tau$ between the two signals. The relevant scale is the y-axis on the left, where we see that biases $\mathcal{R}>1$ can arise. In gray, we indicate the region where we regard biases as not significant ($\mathcal{R}<1$). In warm colors, we plot the correlation coefficients (with relevant y-axis on the right), defined in Eq.~\eqref{eq:corrcoeff}. We see that the largest correlations $\sigma\gtrsim 0.05$ correspond to the largest biases ($\sim 6\sigma$).}
	\label{fig:R_theta_correlations}
\end{figure}

We consider an ET data stream composed of a signal $h^{(1)}$ to be inferred and a missed signal $h^{(2)}$ that creates confusion noise \begin{equation}
    \hat d(f) =\hat h_e^{(1)} (f; \boldsymbol{\theta}^{(1)}) + \hat h_e^{(2)} (f; \boldsymbol{\theta}^{(2)})\,.
\end{equation}
For this example we ignore waveform errors and detector noise. The biases arise solely due to the confusion noise $\hat h_e^{(2)}$, and can be predicted from~\eqref{multipar} setting $n=\delta h=0$. 
The parameter space of the Fisher matrix is $\boldsymbol\theta^{(1)}=\{\log \mathcal{M}_c^{(1)},\eta^{(1)}\}$, with true parameters $\boldsymbol{\theta}^{(1)}_\text{tr}=\{15.4 M_\odot, 0.243\}$ (corresponding to a binary with component masses $m_1= 21M_\odot$ and $m_2 = 15 M_\odot$). We take the signal to be nonspinning ($\beta^{(1)}=0$) with an effective distance $D_\text{eff}^{(1)}=5$Gpc, and phase and times at coalescence $\phi^{(1)}_c = \pi/3$ and $t_c^{(1)}=0s$. The SNR for this source is $\rho(h^{(1)})\sim 75$. For the overlapping signal, we pick component masses $m_1 = 25M_\odot$ and $m_2 = 20M_\odot$, a nonspinning configuration $\beta^{(2)}=0$, an effective distance $D_\text{eff}^{(2)}=10$Gpc, and phase at coalescence $\phi^{(0)}_c = \pi/3$. We let $t_c^{(2)}$ vary as a free parameter. For a nominal value of $t_c^{(2)}=-0.2s$, the SNR for the overlapping source is $\rho(h^{(2)})\sim 46$. In Fig.~\ref{fig:TD_waveforms}, we plot time-domain waveforms for this particular configuration.

We now turn to the problem of predicting the biases on $ \boldsymbol{\theta}^{(1)}$. From Eqs.~\eqref{multipar} and~\eqref{eq:R_func}, we compute the bias ratio $\mathcal{R}(\Delta\boldsymbol\theta_\text{conf}^{(1)})$ due to the presence of confusion noise, varying $\tau :=t^{(2)}_c-t^{(1)}_c$ between $\tau = -2.0$ and $\tau = 2.0$. The results are shown in Fig.~\ref{fig:R_theta_correlations}. In this Figure, we plot both the ratios $\mathcal{R}$ and the Pearson correlation coefficients\footnote{These correlations are calculated using the joint Fisher matrix, which fundamentally assumes that we have resolved \emph{both} signals. In this case, we would expect no biases from the overlapping signal. In the bias ratios calculation, we treat the second signal as unfitted, which leads us to the shown biases from confusion noise. 
Regardless of this difference in treating the Fisher matrix, we conclude that Pearson correlations can be a guide to understand where biases would occur if the overlapping signal were not inferred, as suggested in~\citep{Pizzati:2021gzd}.
}, defined as 
\begin{equation}
    \sigma_{\theta_1 \theta_2} = \frac{(\Gamma^{-1})_{\theta_1 \theta_2}}{\sqrt{(\Gamma^{-1})_{\theta_1 \theta_1}(\Gamma^{-1})_{\theta_2 \theta_2}}}\,.\label{eq:corrcoeff}
\end{equation}
We notice that non-trivial biases start appearing when $|\tau|\lesssim 0.5$, which correspond to the largest correlation coefficients ($\sigma_{\theta_1 \theta_2}\sim 0.05$). We therefore (qualitatively) confirm the main result of~\citep{Pizzati:2021gzd} [and of \citep{Samajdar:2021egv,Himemoto:2021ukb,Relton:2021cax} indirectly]. Notice that because of our choice of data input and parameters, our comparisons with the results of~\citep{Pizzati:2021gzd} can only be qualitative. They consider noise in Advanced LIGO, while we consider ET (picking a noiseless realization in the data stream).
Furthermore, they model their signals with a different approximant (\texttt{IMRPhenomv2}), include detector response functions, sample through masses with different true values and include additional parameters in the analysis, specifically the phase, $\phi_c$, and time, $t_c$, at coalescence, and luminosity distance, $d_L$. 

To check the reliability of our bias predictions, we have also compared them against posteriors from an MCMC run for a configuration with the $\tau$ leading to the largest biases ($\sim 6\sigma$, for the $\tau=-0.2s$ configuration shown in Fig.~\ref{fig:TD_waveforms}): we obtain excellent agreement, at the level of the accuracy shown by the (orange) prediction in Fig.~\ref{fig:conf_noise_bias}. This example illustrates the advantage of our formalism, namely that the biases can be cheaply and reliably predicted. Our formalism will be a valuable tool for extending previous Bayesian analyses into regions of parameter space that are difficult to sample with fully Bayesian techniques.

\subsection{Biases from the inaccurate removal of signals}\label{subsec:biases_innacurate_removal}

\begin{figure}
	\includegraphics[width=\linewidth]{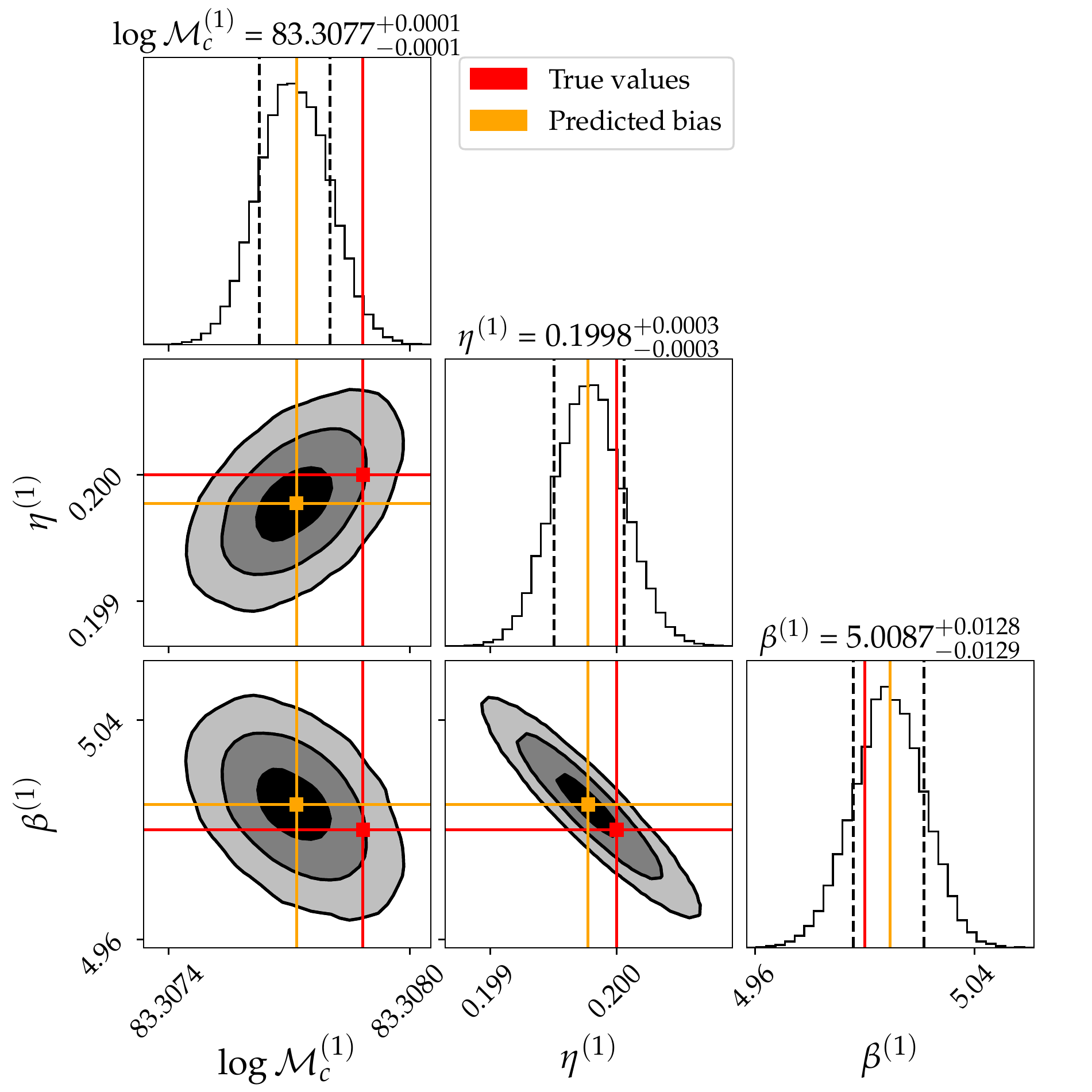}
	\caption{ \emph{Biases from the inaccurate removal of loud sources}. Posterior distributions for the parameters of a reference signal, computed using MCMC, when 2 mismodelled overlapping signals are removed from the data (with parameters given in Table~\ref{tab:example_2}). We also show the biases predicted using our formalism.}
	\label{fig:inacc_rem_bias}
\end{figure}

We now consider the situation in which the confusion sources are not ``missed'', but incorrectly fitted out. To simulate this, we consider a LISA data stream,
\begin{table}
	\centering
	\caption{Parameter configurations for the signals in Sec.~\ref{subsec:biases_innacurate_removal}. We also report the SNR of the source, $\rho_h$ and of the residual $\rho_{\delta h}$. Notice that we do not consider waveform errors for the first (reference) source here, implying its residual is zero. We sample all the sources from $f=0.5$mHz and stop at 2mHz ($T_\text{obs}=0.3$ days), the earliest chirp time for these masses. }
	\label{tab:example_2}
	\begin{tabular}{lccc|ccc|cr} 
		\hline
		i & $M/M_\odot$ & $\eta$ & $\beta$ & $D_\text{eff}$ & $t_c$ & $\phi_c$ & $\rho_h$ & $\rho_{\delta h}$\\
		\hline
		1 & $2\cdot 10^6$  & 0.20 & 5.0 & 10 $\text{Gpc}$& 6 h & 0 & 83 & - \\
		2 & $1\cdot 10^6$  & 0.23 & 1.0 & 3 $\text{Gpc}$& 48 h & $\pi$ & 790 & 31 \\
		3 & $4\cdot 10^6$  & 0.08 & 2.4 & 2 $\text{Gpc}$& 6 h & 0.9 & 2216 & 76 \\
		\hline
	\end{tabular}
\end{table}
\begin{equation}\label{data_inc_rem}
    \hat d(f) = \hat h^{(1)}_e(f; \boldsymbol\theta^{(1)}) + \hat h^{(2)}_e(f; \boldsymbol\theta^{(2)})
    + \hat h^{(3)}_e(f; \boldsymbol\theta^{(3)})
\end{equation}
where the signal ``(1)'' is our reference signal, which we assume is modelled perfectly, and the other sources are incorrectly subtracted using approximate templates $\hat h_m(f; \boldsymbol\theta^{(2,3)},\epsilon=0.3)$. In such a procedure, we expect biases to arise only from the residual that the incorrectly modelled signals leave in the data stream~\eqref{data_inc_rem},
\begin{equation}
\delta h = \sum_{i=2}^{3} \hat h^{(i)}_e(f;\boldsymbol\theta^{(i)})-\hat h^{(i)}_m(f;\boldsymbol\theta^{(i)}, \epsilon=0.3)\,.
\end{equation}
In this case, the  relevant parameter space is $\boldsymbol\Theta =\{\boldsymbol\theta^{(1)},\boldsymbol\theta^{(2)},\boldsymbol\theta^{(3)}\}$, where we pick each subset to be $\boldsymbol\theta^{(i)}=\{\log \mathcal{M}_c^{(i)},\eta^{(i)},\beta^{(i)}\}$. The joint Fisher matrix $\Gamma$ is therefore a 9$\times$9 matrix (calculated using $\hat h_m$). We report the true source parameters in Table~\ref{tab:example_2}. We calculate the biases $\Delta\boldsymbol\theta^{(1)}$ on the reference signal's parameters using~\eqref{eq:source_1_bias}(or equivalently~\eqref{multipar}), which leads us to
\begin{align}\label{R_res_inacc}
    &\mathcal{R}(\Delta\log \mathcal{M}_c^{(1)})=1.98 >1\nonumber\\
    &\mathcal{R}(\Delta\eta^{(1)})=0.84\nonumber\\
    &\mathcal{R}(\Delta\beta^{(1)})=0.74\,.
\end{align}
Biases are then significant for the chirp mass in this case. These predictions can be checked with an MCMC analysis, see Fig.~\ref{fig:inacc_rem_bias}. We find that the formalism can accurately predict the biases from the inaccurate removal of signals. 

The fact that each contribution to $\delta h$ in Eqs.~(\ref{eq:source_1_bias},\ref{multipar}) affects the parameters of each source equally suggests that residuals effectively behave as missed sources and confusion noise. In fact, we can rewrite the data stream analysed in Fig.~\ref{fig:inacc_rem_bias} in the form
\begin{align}\label{data_inc_rem_check}
    \hat d(f) =& \hat h^{(1)}_e(f; \boldsymbol\theta^{(1)}) +\nonumber\\
    &\hat h^{(2)}_m(f; \boldsymbol\theta^{(2)},\epsilon=0.3) + \hat h^{(3)}_m(f; \boldsymbol\theta^{(3)},\epsilon=0.3)+\delta h,
\end{align}
which explicitly separates out the modelled part using the models employed by the MCMC analysis and the calculation of the joint Fisher matrix. Doing so leaves an extra term, $\delta h$, which plays the role of the confusion noise caused by the residuals.
One can check that the biases predicted from the data stream~\eqref{data_inc_rem_check} (and obtained using the joint Fisher matrix with $\hat h_m$) match the predictions reported in Fig.~\ref{fig:inacc_rem_bias}.
An important implication of this equivalence of results is that significant biases may arise from the incorrect removal of a very large number of signals drawn from the same population, in direct analogy with the findings of the previous section. We have not checked this directly, since adding a considerable number of fitted sources dramatically increases the
dimensionality of $\boldsymbol\Theta$, making the implementation of the joint Fisher matrix difficult.

%%%%%%%%%%%%%%%%%%%%%%%%%%%%%%%%%%%%%%%%%%%%%%%%%%
\subsection{Waveform errors \& confusion noise} \label{sec:results}
%%%%%%%%%%%%%%%%%%%%%%%%%%%%%%%%%%%%%%%%%%%%%%%%%%

 \begin{figure*}
     \centering
     \includegraphics[height = 15cm,width = 18cm]{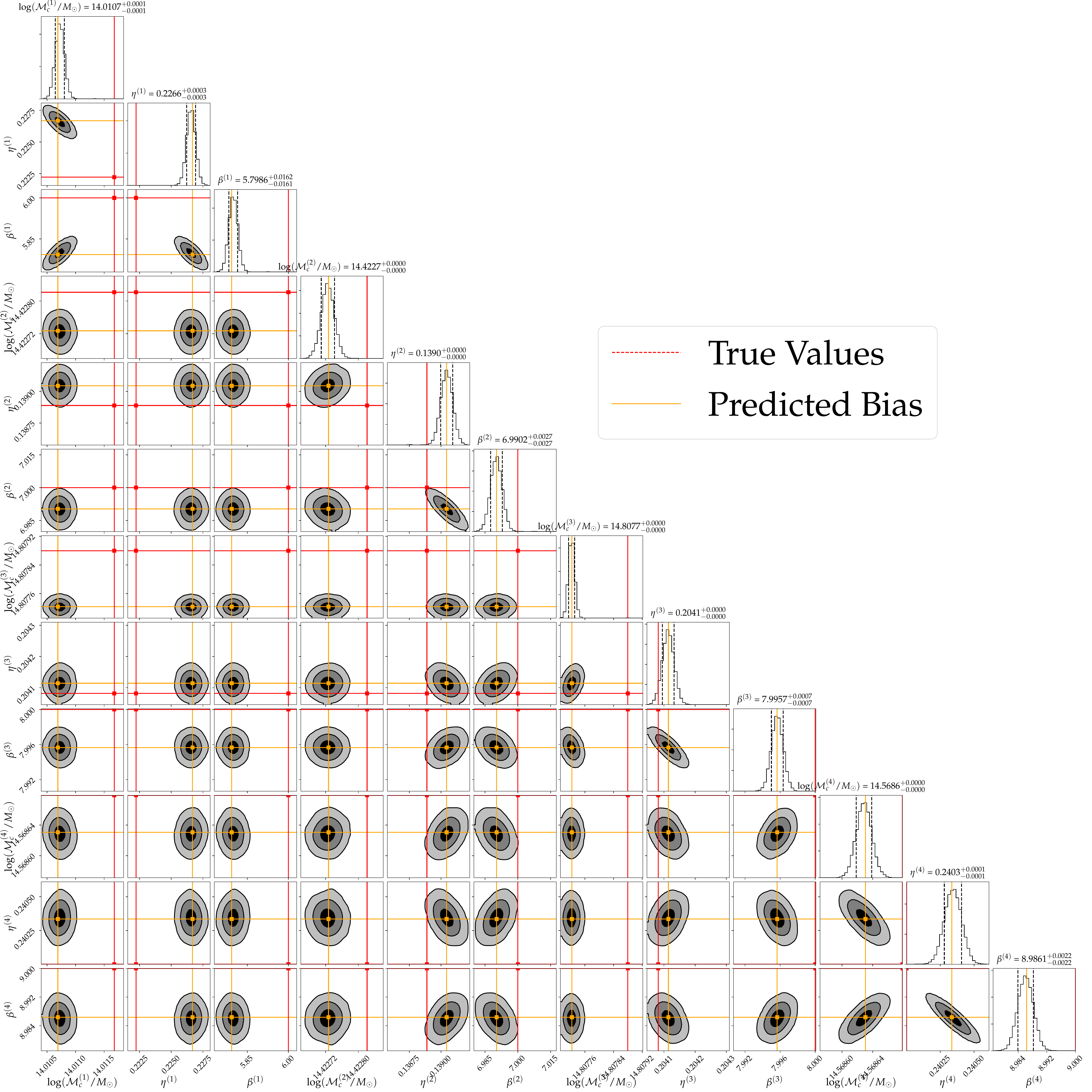}
     \caption{Triangle plot of the one-dimensional (on the diagonal) and two-dimensional marginalised posterior distributions for the inferred parameters in the LISA scenario considered in Sec.~\ref{sec:results}. The red lines indicate the true parameters and orange lines indicate the biases predicted from \eqref{multipar}.}
     \label{fig:LISA_Corner}
 \end{figure*}

\begin{figure*}
    \centering
    \includegraphics[height = 5.5cm, width = 17cm]{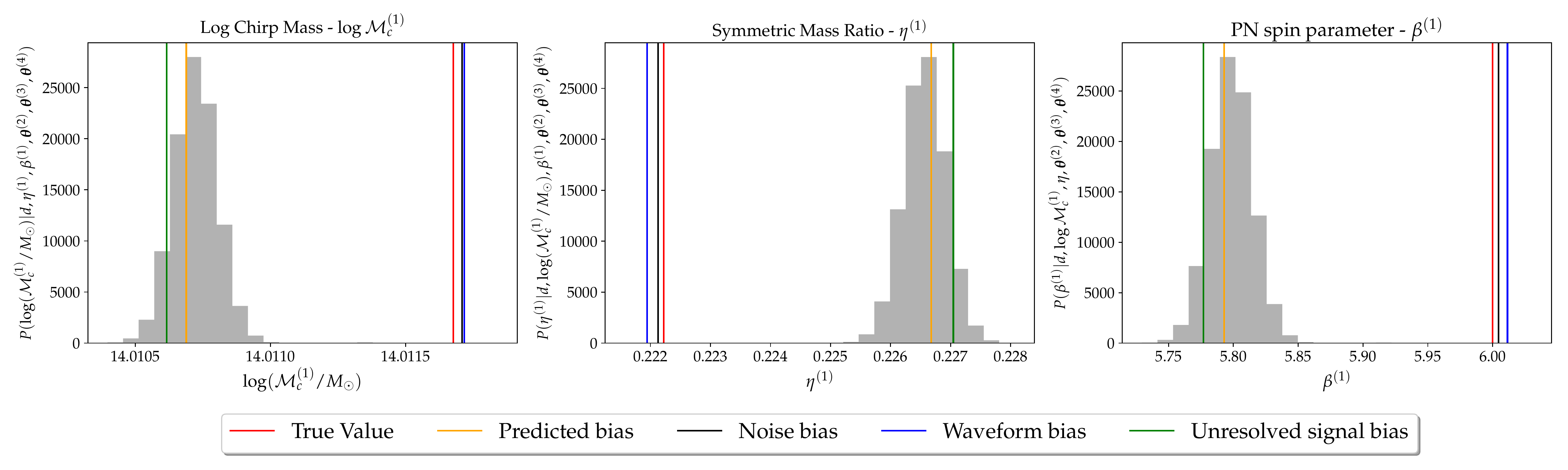}
    \caption{(top/bottom left to right) The grey histograms are the posterior samples for $\log M_{c}^{(1)},\eta^{(1)}$ and $\beta^{(1)}$ for the LISA scenario considered in Sec.~\ref{sec:results}. The red lines indicate the true parameters, blue lines the biases arising from the use of inaccurate waveforms as templates, the black ones the noise induced shift in the peak of the likelihood, the green lines the biases due to unresolved signals and the orange lines show the approximation to the total bias computed from \eqref{multipar}.}
    \label{fig:posteriors_LISA}
\end{figure*}

We now bring together the ideas described in sections (\ref{sec:source_conf}) and (\ref{subsec:biases_innacurate_removal}), and show that the formalism developed in Sec.~(\ref{sec:Source_Confusion_Bias}) can accurately predict biases on parameter estimates when we simultaneously fit $N_{\text{fit}}$ sources with inaccurate waveforms, while confusion and detector noise are also present in the data stream. We show this here for LISA, but an ET example may be found in Appendix~\ref{app:ET_gf}. The data stream in this case is
\begin{align}
    \hat{d}(f) &=  \sum_{i=1}^{N_{\text{fit}}}\hat{h}^{(i)}_{e}(f;\boldsymbol{\theta}^{(i)}_{\text{tr}}) + \Delta H_\text{conf} + \hat{n}(f).\label{eq:global_eq}
\end{align}
We assume $\Delta H_{\text{conf}}$ arises from the galactic foreground of white-dwarf binaries (WDB). LISA is guaranteed to detect WDBs in their thousands or even tens of thousands~\citep{Crowder_2007,B_aut_2010} (depending on the imposed SNR threshold), but there will also be millions of unresolved WDBs radiatig in the LISA band. Here we assume that WDBs with $\rho < 8$ have been folded into the PSD~\citep{B_aut_2010}.
We additionally assume that only WDBs with $\rho > 15$ have been detected by dedicated pipelines, which leaves us with missed WDBs with SNRs in the range $8< \rho< 15$. To simulate these sources, we construct a superposition of signals, see Eq.~\eqref{confres}, with frequencies chosen from $f_{i}\in(10^{-4},10^{-3})$Hz. For simplicity, we only retain the leading PN term in the waveform, computed for random masses drawn from $(m_{1},m_{2}) \sim 10^{2}\cdot U^{2}[0.3,1]M_{\odot}$.

We finally draw effective distances $D_{\text{eff}} \sim 10^{4}\cdot U[1,3]\text{pc}$. We discard binaries not in the specified range of SNRs, until $N_U=1000$ are found.
To complete the input data stream, we add $N_{\text{fit}} =4$ fitted signals with waveform errors $\epsilon = 0.04$ and source parameters $\boldsymbol{\theta}^{(i)}$ given in Tab.\ref{tab:sec_5.3_table}. We choose initial frequencies $f_{0} = 10^{-4}$Hz and sample the sources simultaneously with a maximum frequency given by the highest ISCO frequency among the fitted sources. For simplicity, we set $(\phi_{c},t_{c}) = (0,T_{\text{min}})$ for all sources, where $T_{\text{min}}$ is the minimum chirping time allowed over all parameter configurations. The SNRs are of order $\mathcal{O}(10^{3})$ for all fitted sources. 

\begin{table}
	\centering
	\caption{Parameters for the simultaneously-fitted LISA signals in Sec.~\ref{sec:results}.}
	\label{tab:sec_5.3_table}
\begin{tabular}{c|cccc}
$i$                   & $M/M_{\odot}$ & $\eta$ & $\beta$ & $D_{\text{eff}}/\text{Gpc}$   \\ \hline
1                     & $3 \times 10^{6}$                &  0.222      &  6       & 2    \\
2                     & $6 \times 10^{6}$               &   0.139     &  7     &  3      \\
3                     &    $7 \times 10^{6}$            &   0.204     &    8      & 1   \\
4                     &    $5 \times 10^{6}$            &   0.240     &    9     &  1 
\end{tabular}
\end{table}

Corner plots displaying \emph{all} parameter biases can be found in Fig.~\ref{fig:LISA_Corner}. We see that the predicted biases from~\eqref{multipar} are in remarkable agreement with the posteriors from the MCMC algorithm.  
Additionally, in Fig.\ref{fig:posteriors_LISA} we show how the total shift in the peak of the posterior of the parameters $\boldsymbol{\theta}^{(1)}$ of the first source, computed from Eq.~\eqref{eq:source_1_bias}, breaks down into its constituent contributions. 
Firstly, we see that biases from confusion noise, unresolved sources or waveform residuals can deconstructively interfere, i.e., the combined contribution can be smaller than the worst of the individual contributions. 
Secondly, we notice that there are large biases from confusion noise, which implies that if global-fit analyses miss $\mathcal{O}(1000)$ WDBs, this will lead to a significant bias in parameter estimates for other GW sources. We have further explored how biases change when the threshold is taken to be any value $\rho_\text{th} \in [8,15]$. We have tested that, when this threshold is increased towards $\rho_\text{th}=15$, biases tend to increase as the SNR of $\Delta H_\text{conf}$ increases. While the model used here is approximate, it suggests that the completeness of LISA data analysis algorithms needs to be sufficiently high down to sufficiently low threshold SNRs for biases on other parameters to be minimized.

\section{global-fit schemes}\label{sec:GF}

So far, we have defined the global-fit as the simultaneous search for and parameter estimation of \emph{all} gravitational wave signals in the LISA data stream. In Sec.~\ref{sec:results}, this was achieved by assuming the number of signals (and the associated parameter space) present in the data stream was known precisely. However, in a realistic scenario, we will not know how many signals are present in the data. Furthermore, 
the number of signals present at any given time may be large, leading to a prohibitively large parameter space. Consider, for example, the simultaneous inference of an extreme mass-ratio inspiral (a small compact object inspiraling into a super massive black hole) and a massive black-hole binary. Both systems will have parameter spaces $\gtrsim$ 14 dimensions, requiring parameter estimation algorithms to sample from a $\gtrsim 28$ dimensional posterior. This could stretch the capabilities of current inference techniques (especially when correlations between parameters of different sources are high). The problem is likely to worsen as more signals are included in the model. One solution is to use state-of-the-art parameter estimation techniques that are able to efficiently sample such complicated, high-dimensional posterior distributions. In principle, such methods would be no more computationally expensive than the method we describe here. However, it is likely to be difficult to design an algorithm that can robustly and efficiently sample from the full global-fit posterior, and so it is valuable to consider alternative approaches that are easier to implement, and more robust. We will describe one such alternative idea in this section. %is to provide alternative ideas to tackle the global fit problem. 
We begin by proposing an (expensive) iterative approach to sample reduced portions of the parameter space. Then, using the formalism developed above, we illustrate how to cheaply correct for the biases arising within the first few parameter estimation simulations. The final posterior estimates will not be as accurate as those from a simultaneous global-fit, and so this algorithm cannot fully replace a general global-fit analysis. However, the approach is worth exploring as it could provide a quicker and easier way to obtain an accurate initial estimate of the source parameters and their uncertainties. This could then be used to assist the global-fit, for example by providing a starting point for further sampling and refinement, or by providing a proposal distribution to use within the global-fit sampler, or by just providing a cross-check of the results~\footnote{We note here that cross checks are likely to be useful only in the domain in which the Fisher matrix is a good approximation for all the considered parameters. The range of applicability of the Fisher matrix, whose extent is to be substantiated with future analyses, may be further restricted with the addition of realistic features such as the detector response functions.}, to ensure that the global-fit sampler has converged.

\subsection{Parameter Estimation through local-fits}
Let $h^{(\mathcal{A})} \in \mathcal{A}$ and $h^{(\mathcal{B})} \in \mathcal{B}$ denote a set of distinct signals with parameters $\boldsymbol{\theta}^{\mathcal{A}}_{\text{tr}}$ and $\boldsymbol{\theta}^{\mathcal{B}}_{\text{tr}}$ we wish to infer. The joint data stream is given by 
\begin{equation}\label{eq:gf_joint_data_stream}
    d(\mathcal{A},\mathcal{B}) = \sum_{\mathcal{A}} h^{(\mathcal{A})}_{e}(\boldsymbol{\theta^{\mathcal{A}}_{\text{tr}}}) + \sum_{\mathcal{B}} h^{(\mathcal{B})}_{e}(\boldsymbol{\theta^{\mathcal{B}}_{\text{tr}}}) + n(t).
\end{equation} 
For simplicity, we ignore effects coming from unresolved signals. Global-fit pipelines are concerned with the data stream \eqref{eq:gf_joint_data_stream} with the goal to simultaneously infer both signal sets $h^{(\mathcal{A})} \in \mathcal{A}$ and $h^{(\mathcal{B})} \in \mathcal{B}$.

In a local-fit procedure, we consider performing parameter estimation \emph{only} on signal set $\mathcal{A}$ and treat signals from the set $\mathcal{B}$ as missed signals. 
We write this data stream as
\begin{align}
 d(\mathcal{A}|\mathcal{B}) &= \sum_{\mathcal{A}} h^{(\mathcal{A})}_{e}(\boldsymbol{\theta^{\mathcal{A}}_{\text{tr}}}) + \Delta H_{\text{conf}} + n(t),  \label{eq:gf_A_g_B}\\
 \Delta H_{\text{conf}} &= \sum_{\mathcal{B}} h^{(\mathcal{B})}_{e}(\boldsymbol{\theta^{\mathcal{B}}_{\text{tr}}}).
\end{align}
The best fit parameters for $\mathcal{A}$ obtained in this stage can be denoted $\boldsymbol{\theta}^{\mathcal{A}|\mathcal{B}}_{\text{bf}}$, the conditioning on ${\cal B}$ indicating that the estimate was obtained with ${\cal B}$ present in the data. In the second step, we use the recovered parameters $\boldsymbol{\theta}^{\mathcal{A}|\mathcal{B}}_{\text{bf}}$ to subtract out an estimate of $h^\mathcal{A}$ from the joint data stream using our approximate model
\begin{equation}\label{eq:gf_PE_B_A_step_1}
    d(\mathcal{B}|\mathcal{A}_{\text{res}}) = d(\mathcal{A},\mathcal{B}) - \sum_{\mathcal{A}}h_{m}^{(\mathcal{A})}(\boldsymbol{\theta}^{\mathcal{A}|B}_{\text{bf}}).
\end{equation}
Then one estimates the parameters of signal $\mathcal{B}$ using the data stream \eqref{eq:gf_PE_B_A_step_1} with signal templates representing signals in $\mathcal{B}$. This will yield parameters $\boldsymbol{\theta}^{\mathcal{B}|\mathcal{A}_{\text{res}}}_{\text{bf}}$, where $\mathcal{A}_{\text{res}}$ indicates that this analysis was done on a ``residual data set'' from which an estimate of $h^{\cal A}$ had been subtracted. This estimate can be used to \emph{update} the initial data stream $d(\mathcal{A}|\mathcal{B})$, now denoted $d(\mathcal{A}|\mathcal{B}_{\text{res}}) = d(\mathcal{A},\mathcal{B}) - \sum_{\mathcal{B}}h^{(\mathcal{B})}(\boldsymbol{\theta}^{\mathcal{B}|\mathcal{A}_{\text{res}}}_{\text{bf}})$. Again, we can perform parameter estimation on signals $\mathcal{A}$, now with residuals from $\mathcal{B}$ in the data stream, using this updated data array and recovering $\boldsymbol{\theta}^{\mathcal{A}|\mathcal{B}_{\text{res}}}_{\text{bf}}$. These recovered parameters should be \emph{closer} to the true parameters than $\boldsymbol{\theta}^{\mathcal{A}|\mathcal{B}}_{\text{bf}}$. We can continue this scheme by then searching over \begin{equation}\label{eq:gf_PE_B_A_step_2}
    d(\mathcal{B}|\mathcal{A}_{\text{res}}) = d(\mathcal{A},\mathcal{B}) - \sum_{\mathcal{A}}h_{m}^{(\mathcal{A})}(\boldsymbol{\theta}^{\mathcal{A}|\mathcal{B}_{\text{res}}}_{\text{bf}}),
\end{equation} recovering parameters, then searching over $d(\mathcal{A}|\mathcal{B}_{\text{res}})$, and so on and so forth. What we would find is that the recovered parameters for both $\boldsymbol{\theta}^{\mathcal{A}}_{\text{bf}}$ and $\boldsymbol{\theta}^{\mathcal{B}}_{\text{bf}}$ tend towards the ``true'' parameters, i.e., the parameters that would have been recovered if a global fit procedure was carried out. %This is because residuals in equation \eqref{eq:gf_PE_B_A_step_1} are larger than in equation \eqref{eq:gf_PE_B_A_step_2}. 
An advantage of this procedure is that it sidesteps issues arising from sampling the joint posterior for $\mathcal{A}$ and $\mathcal{B}$, but a clear disadvantage is that it requires a number of repeated parameter inference calculations. Computationally, this is expensive and time consuming. As an alternative, we propose that one can use the algorithm presented in Sec.(\ref{sec:Source_Confusion_Bias}) to correct the biases found above. In doing so, one may be able to get a reliable estimate of the true parameters $\boldsymbol{\theta}^{(\mathcal{A})}_{\text{tr}}$ and $\boldsymbol{\theta}^{(\mathcal{B})}_{\text{tr}}$ without having to iterate, i.e., using just the first two parameter inference calculations.%perform as many parameter estimation simulations.

\subsection{Correcting biases in the local-fit analysis}

Before we talk about the details of our algorithm, it is instructive to discuss the source of the biases in parameters $\boldsymbol{\theta}^{\mathcal{A}|\mathcal{B}}_{\text{bf}}$ and $\boldsymbol{\theta}^{\mathcal{B}|\mathcal{A}_{\text{res}}}_{\text{bf}}$. %These come from both $\delta h$, missed signals $\Delta H_{\text{conf}}$ and detector noise $n(t)$ present in  in Eq.~\eqref{multipar}. 
For the data stream \eqref{eq:gf_A_g_B}, the bias in the recovered parameter $\boldsymbol{\theta}^{\mathcal{A|B}}_{\text{bf}}$ is sourced by
\begin{equation}\label{eq:gf_A_g_B_h}
    \delta h^{\mathcal{A}|\mathcal{B}} = \sum_{\mathcal{B}} \hat{h}^{(\mathcal{B})}_{e}(\boldsymbol{\theta}_{\text{tr}}) +
    \sum_{\mathcal{A}}\left[\hat{h}_{e}^{(\mathcal{A})}(\boldsymbol{\theta}^{(\mathcal{A})}_{\text{tr}}) - %\hat{h}_{m}^{(\mathcal{A})}(\boldsymbol{\theta}^{\mathcal{A}|\mathcal{B}}_{\text{bf}})\right]  + \hat{n}(f),
\hat{h}_{m}^{(\mathcal{A})}(\boldsymbol{\theta}^{(\mathcal{A})}_{\text{tr}})\right]  + \hat{n}(f),
\end{equation}
and similarly the bias in $\boldsymbol{\theta}^{\mathcal{B}}_{\text{tr}}$ when performing PE on the data stream \eqref{eq:gf_PE_B_A_step_1}
\begin{multline}\label{eq:gf_B_g_A_res_h}
    \delta h^{\mathcal{B}|\mathcal{A}_{\text{res}}} = \sum_{\mathcal{A}} \left[\hat{h}_{e}^{(\mathcal{A})}(\boldsymbol{\theta}^{(\mathcal{A})}_{\text{tr}}) - \hat{h}_{m}^{(\mathcal{A})}(\boldsymbol{\theta}^{\mathcal{A}|\mathcal{B}}_{\text{bf}})\right]
    + \\
    \sum_{\mathcal{B}} \left[\hat{h}_{e}^{(\mathcal{B})}(\boldsymbol{\theta}^{(\mathcal{B})}_{\text{tr}}) - 
    %\hat{h}_{m}^{(\mathcal{B})}(\boldsymbol{\theta}^{\mathcal{B}|\mathcal{A}_{\text{res}}}_{\text{bf}})\right]
\hat{h}_{m}^{(\mathcal{B})}(\boldsymbol{\theta}^{(\mathcal{B})}_{\text{tr}})\right]
    + \hat{n}(f).
\end{multline}
In Eq.\eqref{eq:gf_A_g_B_h}, the first term is the bias due to missed signals $\mathcal{B}$, the second term the residuals due to incorrect subtraction of the true signals and finally the noise. The noise related bias should be consistent with the width of the posterior. Also, the errors due to inaccurate waveforms should decrease as more accurate waveforms are developed. Thus, we believe it is reasonable to assume that the dominant contribution to the bias comes from the first term in Eq.\eqref{eq:gf_A_g_B_h}. A similar story can be told for Eq.\eqref{eq:gf_B_g_A_res_h} where we expect the first term will dominate and the latter two will be subdominant corrections.
Finally, we do not have access to the true parameters $\boldsymbol{\theta}^{\mathcal{A}}_{\text{tr}}$ and $\boldsymbol{\theta}^{\mathcal{B}}_{\text{tr}}$, nor the exact models for $h_{e}^{(\mathcal{A})}$ or $h_{e}^{(\mathcal{B})}$. We make a further approximation for the $\mathcal{B}$ true parameters $\boldsymbol{\theta}^{\mathcal{B}}_{\text{tr}} \approx  \boldsymbol{\theta}^{\mathcal{B}|\mathcal{A}_{\text{res}}}_{\text{bf}}$ and assume that $h_{e} \approx h_{m}$. We have access to these parameters from our first parameter estimation run on signal set  $\mathcal{B}$ using the data stream $d(\mathcal{B}|\mathcal{A}_{\text{res}})$. From this information, we can approximate both Eqs.\eqref{eq:gf_A_g_B_h} and Eq.\eqref{eq:gf_B_g_A_res_h} by 
\begin{align}
    \delta h^{\mathcal{A}|\mathcal{B}} &\approx \sum_{\mathcal{B}} \hat{h}^{(\mathcal{B})}_{m}(\boldsymbol{\theta}^{\mathcal{B}|\mathcal{A}_{\text{res}}}_{\text{bf}}) \label{eq:gf_A_B_approximate}\\
    \delta h^{\mathcal{B}|\mathcal{A}_{\text{res}}} &\approx \sum_{\mathcal{A}} \left[\hat{h}_{m}^{(\mathcal{A})}(\boldsymbol{\theta}^{(\mathcal{A})}_{\text{tr}}) - \hat{h}_{m}^{(\mathcal{A})}(\boldsymbol{\theta}^{\mathcal{A}|\mathcal{B}}_{\text{bf}})\right]. \label{eq:gf_B_A_res_approximate}
\end{align}
A similar complication arises from our lack of access to $\boldsymbol{\theta}^{\mathcal{A}}_{\text{tr}}$ in Eq.\eqref{eq:gf_B_A_res_approximate}. However, the true parameter $\boldsymbol{\theta}^{\mathcal{A}}_{\text{tr}}$ can be estimated by calculating the CV bias using $\delta h^{\mathcal{A}|\mathcal{B}}$ from Eq.\eqref{eq:gf_A_B_approximate} with the Fisher matrix and numerical derivatives calculated at parameter values $\boldsymbol{\theta}^{\mathcal{A}|\mathcal{B}}_{\text{bf}}$. This will produce an estimate of the bias, $\Delta\boldsymbol{\theta}^{\mathcal{A}|\mathcal{B}}$, which can be  subtracted from $\boldsymbol{\theta}^{\mathcal{A}|\mathcal{B}}_{\text{bf}}$, to give an updated estimate of $\boldsymbol{\theta}^{\mathcal{A}}$ that should lie closer to the true parameters, $\boldsymbol{\theta}^{\mathcal{A}}_{\text{tr}}$. This new parameter $\widehat{\boldsymbol{\theta}^{\mathcal{A}|\mathcal{B}}_{\text{bf}}} =  \boldsymbol{\theta}^{\mathcal{A}|\mathcal{B}}_{\text{bf}} - \Delta \boldsymbol{\theta}^{\mathcal{A}|\mathcal{B}}$ can be used to approximate $\boldsymbol{\theta}^{\mathcal{A}}_{\text{tr}}$ in Eq.\eqref{eq:gf_B_A_res_approximate}. Finally, using parameter values $\boldsymbol{\theta}^{\mathcal{B}|\mathcal{A}}_{\text{res}}$ to evaluate waveform derivatives and Fisher matrices, one can compute a new estimate of the bias in the $\mathcal{B}$ set signal parameters, $\Delta\boldsymbol{\theta}^{\mathcal{B}|\mathcal{A}_{\text{res}}}$ by using Eq.\eqref{eq:gf_B_A_res_approximate} with $\widehat{\boldsymbol{\theta}^{\mathcal{A}|\mathcal{B}}_{\text{bf}}} \approx \boldsymbol{\theta}^{A}_{\text{tr}}.$ This new bias can be used to update our best guess for the true parameters if the set of $\mathcal{B}$ signals, namely $\widehat{\boldsymbol{\theta}^{\mathcal{B}|\mathcal{A}_{\text{res}}}_{\text{bf}}} =  \boldsymbol{\theta}^{\mathcal{B}|\mathcal{A}_{\text{res}}}_{\text{bf}} - \Delta \boldsymbol{\theta}^{\mathcal{B}|\mathcal{A}_{\text{res}}}$. By construction, the parameter values $\widehat{\boldsymbol{\theta}^{\mathcal{A}|\mathcal{B}}_{\text{bf}}}$ and  $\widehat{\boldsymbol{\theta}^{\mathcal{B}|\mathcal{A}_{\text{res}}}_{\text{bf}}}$ should lie closer to the true values $\boldsymbol{\theta}^{\mathcal{A}}_{\text{tr}}$ and $\boldsymbol{\theta}^{\mathcal{B}}_{\text{tr}}$ respectively. 

To summarise, the algorithm is as follows
\begin{enumerate}
    \item Calculate $\boldsymbol{\theta}^{\mathcal{A}|\mathcal{B}}_{\text{bf}}$ and $\boldsymbol{\theta}^{\mathcal{B}|\mathcal{A}_{\text{res}}}_{\text{bf}}$ by performing PE on signals $\mathcal{A}$ and $\mathcal{B}$ using data streams $d(\mathcal{A}|\mathcal{B})$ then $d(\mathcal{B}|\mathcal{A}_{\text{res}})$.
    \item Calculate 
    \begin{equation}
        \delta h^{\mathcal{A}|\mathcal{B}}_{\text{conf}} \approx \sum_{\mathcal{B}} \hat{h}_{m}(\boldsymbol{\theta}^{\mathcal{B}|\mathcal{A}_{\text{res}}}_{\text{bf}})
    \end{equation}
    and then compute an estimate of the bias on the parameters specific to $\mathcal{A}$, %|\mathcal{B}$, 
    denoted $\Delta \boldsymbol{\theta}^{\mathcal{A}|\mathcal{B}}_{\text{bf}}$, evaluating the waveform derivatives at the parameter values $\boldsymbol{\theta}^{\mathcal{A}|\mathcal{B}}_{\text{bf}}$. Set new best fit parameters for $\mathcal{A}$ %|\mathcal{B}$ 
    as   $\widehat{\boldsymbol{\theta}^{\mathcal{A}|\mathcal{B}}_{\text{bf}}} =  \boldsymbol{\theta}^{\mathcal{A}|\mathcal{B}}_{\text{bf}} - \Delta \boldsymbol{\theta}^{\mathcal{A}|\mathcal{B}}_{\text{bf}}.$
    \item Then calculate
        \begin{equation}
        \delta h^{\mathcal{B}|\mathcal{A}_{\text{res}}}_{\text{conf}} = \hat{h}_{m}(\widehat{\boldsymbol{\theta}^{\mathcal{A}|\mathcal{B}_{\text{res}}}_{\text{bf}}}) -  \hat{h}_{m}(\boldsymbol{\theta}^{\mathcal{A}|\mathcal{B}_{\text{res}}}_{\text{bf}}) 
    \end{equation}
    and calculate the CV bias $\Delta \boldsymbol{\theta}^{\mathcal{B}|\mathcal{A}_{\text{res}}}_{\text{bf}}$ on parameters specific to $\mathcal{B}$ %|\mathcal{A}_{\text{res}}$ 
    using parameter values $\boldsymbol{\theta}^{\mathcal{B}|\mathcal{A}_{\text{res}}}_{\text{bf}}$. Now set new parameters  $\widehat{\boldsymbol{\theta}^{\mathcal{B}|\mathcal{A}_{\text{res}}}_{\text{bf}}} =  \boldsymbol{\theta}^{\mathcal{B}|\mathcal{A}_{\text{res}}}_{\text{bf}} - \Delta \boldsymbol{\theta}^{\mathcal{B}|\mathcal{A}_{\text{res}}}_{\text{bf}}$.
\end{enumerate}

We illustrate the algorithm above by considering a noisy data stream containing two signals, each of which have waveform errors $\epsilon \neq 0$. We lose no generality here since the algorithm presented above is easily generalised to handle a greater number of signals. Thus we consider
\begin{equation}\label{eq:GF_full_data}
\hat{d}(f) = \underbrace{\hat{h}^{(1)}_{e}(f;\boldsymbol{\theta}^{(1)},\epsilon = 10^{-3})}_{\mathcal{A}} + \underbrace{\hat{h}^{(2)}_{e}(f;\boldsymbol{\theta}^{(2)},\epsilon = 10^{-3})}_{\mathcal{B}} + \hat{n}(f).
\end{equation}
With parameters for the $\mathcal{A}$ and $\mathcal{B}$ sources given in table \ref{tab:GF_params}. The results of applying the local-fit procedure are presented in the next section.
\begin{table}
	\centering
	\caption{This table presents the true parameter values for source 1 $(\mathcal{A})$ and source 2 $(\mathcal{B})$ for the example of the local-fit procedure presented in section~\ref{sec:LFres}. %in the form $\{ M,\eta,\beta,D_{\text{eff}}\}$. 
	The SNR of each signal within the data stream $\rho_{h}^2 = (h_{\text{e}}|h_{\text{e}})$ is given in the final column.}
	\label{tab:GF_params}
	\begin{tabular}{c|cccc|c} 
		\hline
		$i$ & $M/M_{\odot}$ & $\eta$ & $\beta$ & $D_\text{eff}/\text{Gpc}$ & $\rho_{h}$ \\
		\hline
		1 $(\mathcal{A})$ & $1.2 \times 10^{7}$  & 0.222 & 8 & 2 & $\sim1850$   \\
		2 $(\mathcal{B})$ & $5 \times 10^{6}$ & 0.160 & 7 & 4 & $\sim379$\\
		\hline
	\end{tabular}
\end{table}
\begin{figure*}
    \centering
    \includegraphics[height = 13cm, width = 17cm]{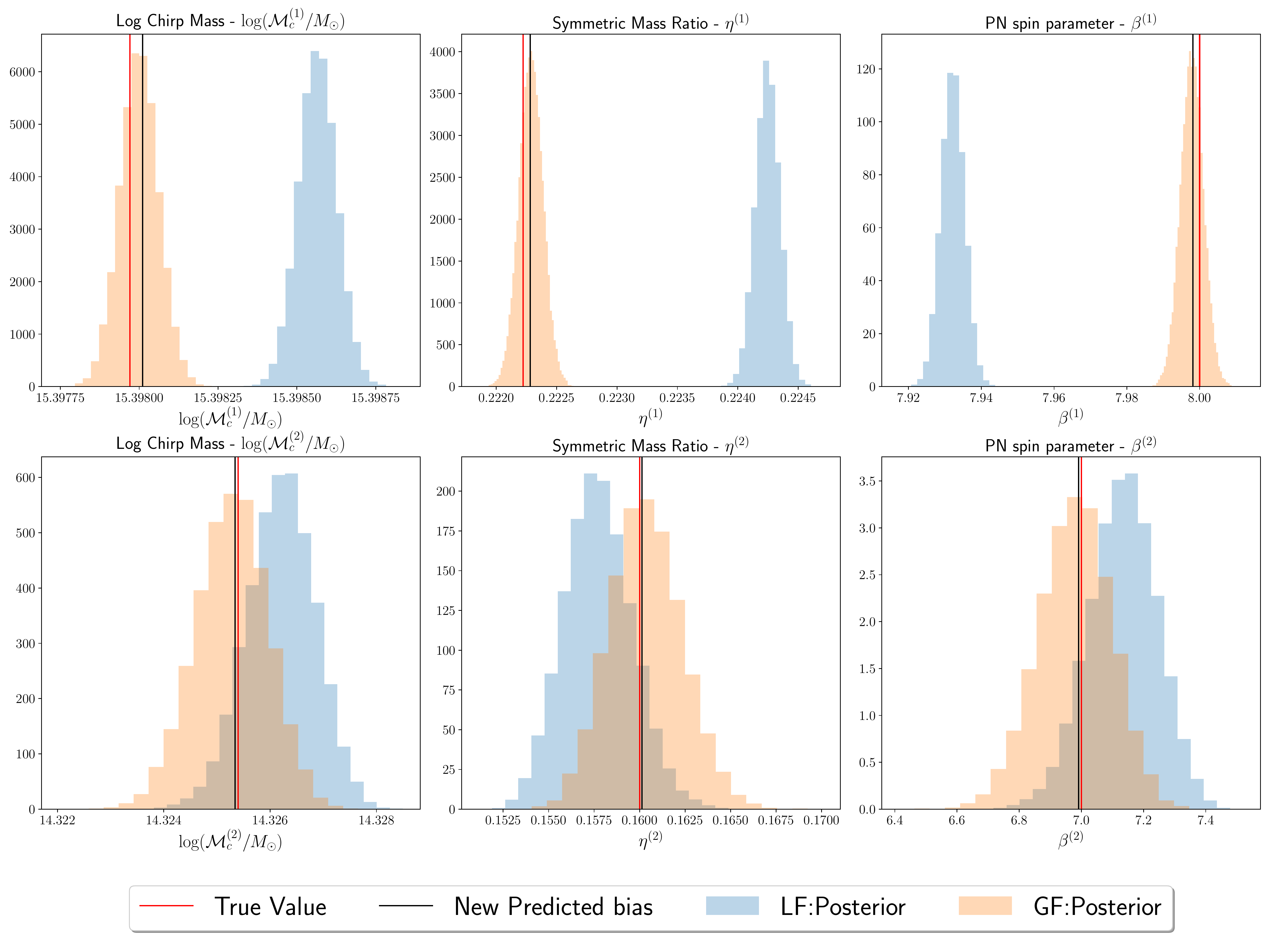}
    \caption{The orange histograms are the global-fit (GF) posteriors from searching the joint data stream $d(\mathcal{A},\mathcal{B})$ for both $\mathcal{A}$ and $\mathcal{B}$ simultaneously. The red lines are true values and black lines the (corrected) new predicted bias using the generalised CV formalism in section \ref{sec:Source_Confusion_Bias}. The blue histograms in the top row are posterior samples from the local-fit (LF) $p(\mathcal{A}|d(\mathcal{A}|\mathcal{B}),\mathcal{B})$ for missed signals $\mathcal{B}$. Similarly, the blue histograms in the bottom row are samples from $p(\mathcal{B}|d(\mathcal{B}|\mathcal{A}_{\text{res}}),\mathcal{A}_{\text{res}})$ for signal $\mathcal{B}$.}
    \label{fig:GF_analysis}
\end{figure*}

\subsection{Results}
\label{sec:LFres}
Following the algorithm above, we present results for the marginalised posteriors in Fig.~\ref{fig:GF_analysis}. In the top row, the blue histogram is the posterior $p(\mathcal{A}|d(\mathcal{A}|\mathcal{B}),\mathcal{B})$ obtained fitting for source $\mathcal{A}$ with source $\mathcal{B}$ in the data, the orange histogram is the posterior for the global-fit solution $p(\mathcal{A},\mathcal{B}|d(\mathcal{A},\mathcal{B}))$, the red lines mark the true parameters and the black line the predicted bias using the formalism. The bottom row of figure \ref{fig:GF_analysis} show corresponding results for the inference of source $\mathcal{B}$, with, for example, the orange histograms representing posterior samples from $p(\mathcal{B}|d(\mathcal{B}|\mathcal{A}_{\text{res}}),\mathcal{A}_{\text{res}})$. In each case, the algorithm is able to correct the bias from the poorly subtracted other signal in the data.
In all cases, after subtracting the predicted bias, the true parameters lie within the $1\sigma$ width of the %global-fit 
posteriors.

In fig.(\ref{fig:GF_analysis}), the local fit posterior for source $\mathcal{B}$ appears to provide a more conservative estimate on how well we can constrain each parameter in comparison to the global fit analysis. Shifting the posterior by the amount predicted by the preceding algorithm will therefore yield a posterior that is broader, and hence more conservative than that which would be obtained from a full analysis. We are yet to develop a strategy to correct parameter uncertainties from the prior local fit analysis. This implies that one must retain precision measurement statements on parameters from the first two parameter estimation runs on $d(\mathcal{A}|\mathcal{B})$ and $d(\mathcal{B}|\mathcal{A}_{\text{res}}).$ Correcting the widths of the local fit posteriors are beyond the scope of this paper and we leave this for future work.

To conclude this section, we make a few important remarks about the algorithm given above. First of all, the algorithm is likely to be less effective if the recovered best fit parameters are far from the true value. This would cause a breakdown of the linear-signal approximation, which is a key assumption in the generalised CV algorithm presented in \ref{sec:Source_Confusion_Bias}. We also assume that, through many local-fits, we have found all the signals present in the data stream we are studying. Further, the two signals present here are near orthogonal with relatively little correlation between the two signals. If there were significant overlap, then the posteriors for the global-fit procedure would be wider since extra uncertainty would be introduced into the parameters in question. 
This would mean that the procedure presented here, in which we shift a posterior computed with a single source model into the correct location, but do not modify the posterior width, would yield overly optimistic estimates of the source parameters. There are two approaches to address this shortcoming. Firstly, the correlation between sources identified in the data can be evaluated, and any pairs of source with sufficiently high correlation can be reanalysed jointly. Alternatively, it is possible to generate an updated posterior for the parameters of each source by marginalising over the biases due to the other source. The procedure is similar to the algorithm described here, but rather than shift each sample in the $\mathcal{A}$ source posterior by the same amount, given by the best-fit parameters of source $\mathcal{B}$, we instead shift them by an amount given by Eq.~\eqref{eq:explainbias} evaluated for the $\mathbf{h}^{(2)}$ waveform computed as a random sample drawn from the $\mathcal{B}$ source distribution. This approach is beyond the scope of the analysis presented here, but we leave it for future work.

\section{Conclusions}\label{conclusions}

%%%%%%%%%%%%%%%%%%%%%%%%%%%%%%%%%%%%%%%%%%%%%%%%%%

In this paper, we have generalized the approach in \citep{Cutler:2007mi} to provide metrics for the parameter estimation biases on individually resolved sources from the presence of confusion noise from missed signals or incorrectly fitted waveforms. We have illustrated these generalisations with simple (yet realistic) scenarios relevant to the LISA and ET detectors, and we can collect several generic findings:

\begin{itemize}
    \item We find that the presence of altogether missed signals drawn from the same population could lead to significant biases on the parameter estimation of other signals which are instead fitted out of the data. 
    \item We qualitatively confirm one of the main results of~\citep{Samajdar:2021egv,Pizzati:2021gzd,Himemoto:2021ukb,Relton:2021cax}. The coincident arrival of two signals in a ground-based detector, with nearly overlapping mergers, may lead to biases when the difference between coalescence times of the signals is less than a fraction of a second.
    \item We find that residuals in the data arising from the incorrect removal of sources effectively behave like missed signals, and may lead to significant biases.
    \item We find that biases from confusion noise and waveform inaccuracies may deconstructively interfere with one another.
    \item Our results suggest that galactic binaries which are missed by dedicated searches~\citep{Littenberg:2020bxy}, and not accounted for in confusion noise estimates, may lead to significant biases on the parameter estimation of other typical LISA sources.
    \item We proposed a proof-of-concept global-fit scheme in which, starting from local-fits of LISA sources, guesses for the true parameters are obtained through bias predictions from previous parameter estimation simulations. We find these guesses lie within the $1\sigma$ interval of global-fit posteriors across all sources. This has potential applications to confirm global-fit search algorithms, and as a standalone novel local-fit parameter estimation algorithm.
\end{itemize}
In all the cases outlined above, the formalism we have developed plays an important role in providing a theoretical ground for the described biases and a solid tool to address them. We believe this formalism could be useful in exploratory studies of future GW detectors, to assess under what circumstances we expect the biases described above to appear.
We also believe this formalism is an early but significant step towards an understanding of how to simultaneously infer parameters from multiple signals of different nature with future detectors, as we highlight with our global-fit algorithm scheme. %\ob{This is good}

There are several ways in which the application of this formalism could be extended. One could perform systematics studies for realistic populations of missed signals using realistically modelled waveforms. One could check whether inaccurately modelled signals could lead to significant biases when several of them are incorrectly subtracted from the data, which our understanding of residuals as missed signals and the biases they lead to strongly suggests. This is a possibility that we have not explored due to the technical challenge in dealing with very large Fisher matrices and MCMC sampling algorithms to sample over such a large parameter space. 
Finally, one could explore further the applications of this formalism for global-fit algorithms, which could be extended to take into account significant overlaps between the signals in the data stream, and to explore correcting the width as well as the peak location of the parameter posteriors.

As a final note, the formalism itself can be extended to take into account brighter confusion sources and more pronounced waveform errors (as would happen with different families of waveform models or models within the same family containing different physics). To do so, one could derive higher order terms in the equations present in Sec.\ref{sec:Source_Confusion_Bias} to describe biases that are farther from the true parameters than those considered in this work. 
\newline

\noindent 
\textit{Acknowledgements.}
The authors thank E.~Berti, D.~Gerosa, M.~P\"urrer,  N.~Tamanini and M.~van de Meent for enlightening discussions. We especially thank R.~Cotesta, M.~Katz and L.~Speri for a careful reading of the manuscript, and R.~Cotesta for collaborating in the early stage of this project as well. The author O.B expresses his gratitude to Sir E. H. John for the vocal support given throughout this work.\\
\noindent
\textit{Data Availability Statement.}
 The data underlying this article will be shared on reasonable request to the corresponding author. Antonelli's and Burke's codes relevant to this project can be found at \url{https://github.com/aantonelli94/GWOP} and 
\url{https://github.com/OllieBurke/Noisy_Neighbours}.

\bibliographystyle{mn2e}
\bibliography{refs}

\appendix
\section{Geometrical interpretation of parameter errors}\label{app:geometry}

\begin{figure}
	\includegraphics[width=\linewidth]{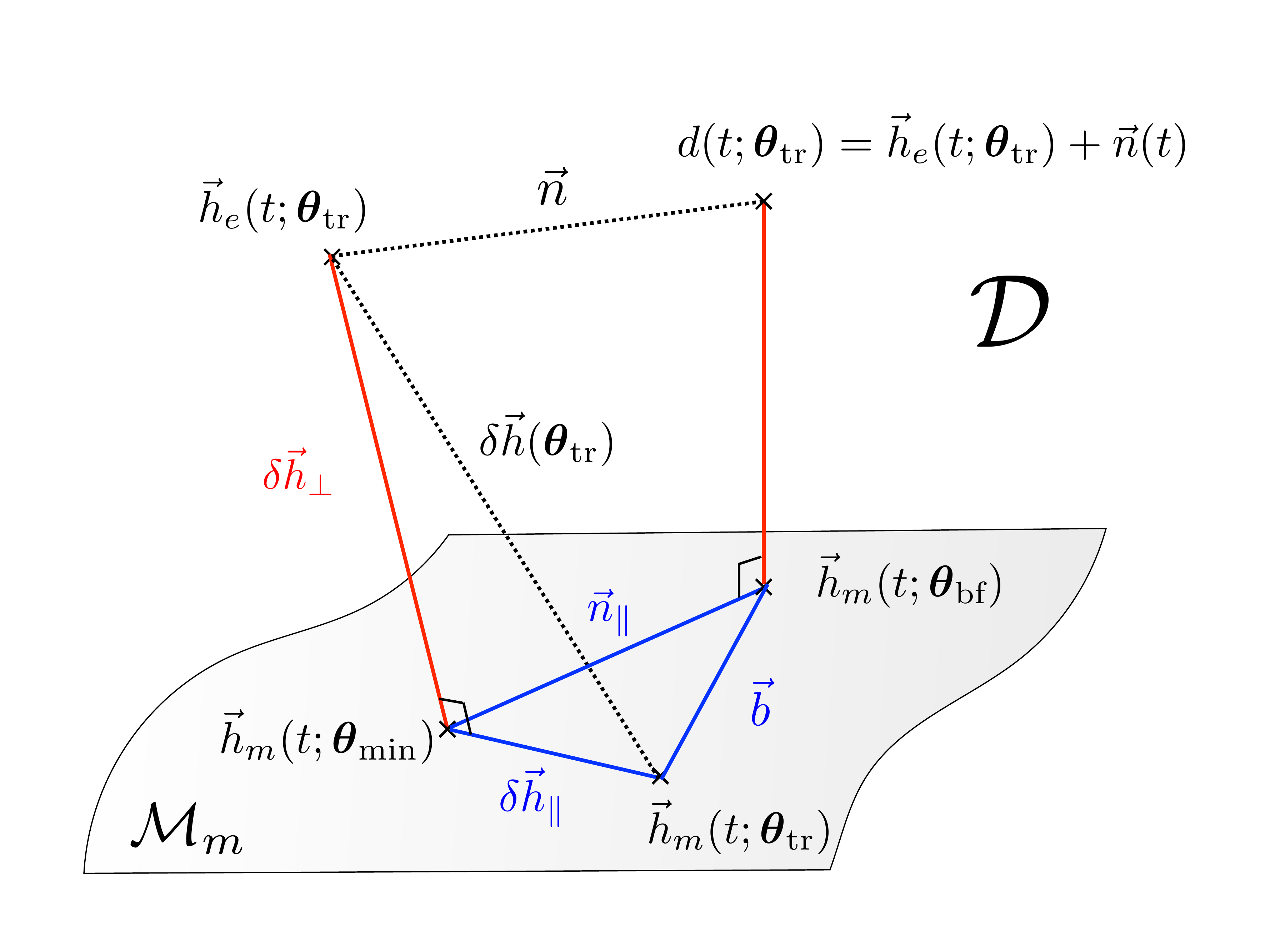}
	\caption{
	Geometrical setup for the CV biases. Represented is the space of signals $\mathcal{D}$ and the various realisations of model and exact templates, with definitions for the parameters as given in the main text. In red, the perpendicular contributions from waveform mismodelling and noise realisations that affect the detectability of the signal. In blue, the contributions to the shifts to the parameters, corresponding to noise-induced errors ($\vec n_\parallel$) and theoretical biases ($\delta\vec h_\parallel$).}
	\label{fig:geometry}
\end{figure}

In this section, we provide a geometrical interpretation for the noise and systematic biases derived in ~\citep{Cutler:2007mi}.
Consider the vector space $\mathcal{D}$ of outputs $\vec d(t;\boldsymbol\theta)$ depending on parameters $\theta^i \in \boldsymbol{\theta}$. Further define two submanifolds $\mathcal{M}_m$ and $\mathcal{M}_e$ of model $\vec h_m(t;\boldsymbol{\theta})$ and fiducial $\vec h_e(t;\boldsymbol{\theta})$ templates, representing both the limiting case of no instrumental noise.
Next, consider the waveform difference $\delta \vec h(\boldsymbol{\theta_\text{tr}}):= \vec h_e(t;\boldsymbol{\theta}_\text{tr})-\vec h_m(t;\boldsymbol{\theta}_\text{tr})$ evaluated at the true parameters.
This can be split into a perpendicular and parallel component. The former is obtained drawing a perpendicular vector $\delta \vec h_\perp$ from $\vec h_e(t;\boldsymbol{\theta}_\text{tr})\in \mathcal{D}$ onto $\mathcal{M}_m$. The projection point is $\vec h_m(t;\boldsymbol\theta_\text{min})$, evaluated at the parameters $\boldsymbol\theta_\text{min}$ that minimise the distance $(\vec h_e - \vec h_m |\vec h_e - \vec h_m)$. Starting from $\vec h_m(t;\boldsymbol\theta_\text{min})$, one can perform a coordinate transformation that maps the model waveform evaluated at $\boldsymbol\theta_\text{min}$ to the same model evaluated at the true parameters $\boldsymbol{\theta}_\text{tr}$. This defines the component $\delta\vec h_\parallel$, see Fig.~(\ref{fig:geometry}). Physically, the $\delta \vec h_\perp$ component corresponds to a ``loss'' of SNR that changes the distance (and therefore affects the likelihood and detectability of the signal only), whereas $\delta h_\parallel \approx (\theta^i_\text{tr} - \theta^i_\text{min}) \partial_i h_m$ corresponds to shifts in the parameters. In what follows, we restrict our attention to vectors in $\mathcal{M}_m$ signalling errors and biases in the parameters, leaving out perpendicular components related to the detectability of the source. 

In a realistic situation, we are confronted with a detector output $\vec d(t; \boldsymbol{\theta}_\text{tr}) = \vec h_e(t; \boldsymbol{\theta}_\text{tr}) +\vec n(t)$ that includes noise. We can project $\vec d$ onto $\mathcal{M}_m$, which defines the model template $\vec h_m(t; \boldsymbol{\theta}_\text{bf})$ evaluated at the best-fit parameters. These are the ones one obtains minimising the argument of the Whittle likelihood, $(\vec d- \vec h_m| \vec d- \vec h_m)$. The new element of $\mathcal{M}_m$, $\vec h_m(t; \boldsymbol{\theta}_\text{bf})$ is connected to  $\vec h_m(t; \boldsymbol{\theta}_\text{min})$ through the parallel component of the noise $\vec n_\parallel$, which can be rewritten as $n_\parallel\approx (\theta^i_\text{bf} - \theta^i_\text{min}) \partial_i h_m$, and to $\vec h_m(t; \boldsymbol{\theta}_\text{tr})$ through a (bias) vector $b\approx (\theta^i_\text{bf} - \theta^i_\text{tr}) \partial_i h_m$, see Fig.~(\ref{fig:geometry}). Then, in this realistic situation the total bias on the PE performed with the model template $\vec h_m$ is given by $\Delta\theta^i:= \theta^i_\text{bf} - \theta^i_\text{tr}$, which itself is formed by two contributions $\Delta\theta^i_\text{noise}:= \theta^i_\text{bf} - \theta^i_\text{min}$ and $\Delta\theta^i_\text{sys}:= \theta^i_\text{tr} - \theta^i_\text{min}$. The former is a statistical error from the noise vector (which averages to zero after many draws of $\vec n$), and we identify it with Eq.~\eqref{eq:width_noise}. The latter is a contribution from waveform mismodelling ($\delta\vec h$) that does not average to zero after many repetitions of the experiment,
and we identify it with the CV bias from theoretical errors~\eqref{eq:width_sys}.

\section{Confusion noise: Stationary treatment}\label{App:confusion_noise}

%\sout{As a final remark we consider the case in which }
When the confusion noise is generated by  a very large population of sources, it is common to treat it analogously to the instrumental noise with $f,f'>0$,
\begin{subequations}
\begin{align}
\langle \widehat{\Delta H_{\text{conf}}} (f) \rangle &= 0, \\
 \langle \widehat{\Delta H_{\text{conf}}} (f) \widehat{\Delta H^{\star}_{\text{conf}}} (f') \rangle &=  \frac{1}{2} \delta(f-f') S_{\text{conf}}(f), 
 \label{eq:confPSD} \\
 \langle \widehat{\Delta H_{\text{conf}}} (f) \widehat{\Delta H_{\text{conf}}} (f') \rangle &=  0 \label{eq:confPSD_no_conj}
\end{align}
\end{subequations}
For $S_{\text{conf}}(f)$ the PSD representing the power of the confusion noise at a particular bin of frequency. In this current discussion we are assuming that the confusion noise acts as a \emph{stationary} time-series that is then fully described by an auto-correlation function.

Under these assumptions, the mean bias is zero and the covariance from the confusion background  
takes the alternative form~\footnote{Note that $S_{\text{conf}}$ describes the contribution from the whole astrophysical population, while $\Sigma_{\text{conf}}$ defined in Eq.~\eqref{eq:sigma_conf} was the contribution from a single source in the population. For consistency, we therefore denote the total covariance by $N \Sigma_{\text{conf}}$ in Eq.~\eqref{eq:sigma_conf_PSD}.}

\begin{align}
    &N\Sigma_{\text{conf}}^{ij} = (\Gamma^{-1})^{ik} (\Gamma^{-1})^{jl} \nonumber \\
    &\hspace{0.4cm} \left\langle \int_{-\infty}^\infty \frac{(\partial_k^* h_m (f) \widehat{\Delta H}(f) + \partial_k h_m (f) \widehat{\Delta H}^*(f))}{S_n(f)} \, {\rm d} f \right. \nonumber \\
    & \hspace{0.4cm}\left. \int_{-\infty}^\infty \frac{(\partial_l^* h_m (f') \widehat{\Delta H}(f') + \partial_l h_m (f') \widehat{\Delta H}^*(f'))}{S_n(f')} \, {\rm d} f'
    \right\rangle \nonumber \\
    &= (\Gamma^{-1})^{ik} (\Gamma^{-1})^{jl} \nonumber \\ 
    &\hspace{0.5cm}2\int_{0}^\infty \frac{(\partial_k h_m^{\star} (f) \partial_l h_m (f) + \partial_l h_m^{\star} (f) \partial_k h_m (f)) S_{\text{conf}}(f)}{S_n^2(f)} \, {\rm d} f
    ,\label{eq:sigma_conf_PSD}
\end{align}
Where we have used \eqref{eq:confPSD}-\eqref{eq:confPSD_no_conj} to reach the final equality.
If we use this prescription within the formalism we have here described, we can calculate the total covariance in the parameter estimates arising from instrumental noise and source confusion, which is $\langle (\Delta \theta^i_{\text{noise}} + \Delta \theta^i_{\text{conf}}) (\Delta \theta^j_{\text{noise}} + \Delta \theta^j_{\text{conf}}) \rangle = \Gamma^{-1} + N\Sigma_{\text{conf}}$, with $\Sigma_{\rm conf}$ defined by Eq.~\eqref{eq:sigma_conf}. This results follows because $\langle \Delta \theta^i_{\text{noise}} \Delta \theta^i_{\text{conf}} \rangle=0,$ since the instrumental and astrophysical noises should not depend on one another. 
To calculate the total variance $[\Gamma^{-1} + \Sigma_{\text{conf}}]^{ij}$, we first quote the general result
\begin{equation}\label{CV_paper:step_1}
\langle (\partial_{i}h_{m}|\widehat{\Delta H_{\text{conf}}})(\partial_{j}h_{m}|\widehat{\Delta H_{\text{conf}}}) \rangle = \Gamma_{ij},
\end{equation}
that is easily proved using \eqref{eq:confPSD}-\eqref{eq:confPSD_no_conj}. We can then re-write $(\Gamma^{-1})^{ij}$ as
\begin{align}\label{CV_paper:step_2}
    (\Gamma^{-1})^{ij} & = \int (\Gamma^{-1})^{ip}\Gamma_{pm}(\Gamma^{-1})^{mj}p_{\text{pop}}(\boldsymbol{\theta}_{\text{conf}})d\boldsymbol{\theta}_{\text{conf}}
\end{align}
since the Fisher matrix is independent of the confusion population and thus population parameters. Integrating over this ensemble of sources is equivalent to taking an ensemble average. Using \eqref{CV_paper:step_1}, \eqref{CV_paper:step_2} and ~\eqref{eq:sigma_conf}, we find 
\begin{align}
   \left[\Gamma^{-1} + \Sigma_{\text{conf}}\right]^{ij} &= (\Gamma^{-1})^{ik}\Sigma_{\text{mix}}^{kl}  (\Gamma^{-1})^{jl},
   \label{eq:biasconfvar}
\end{align}
where
\begin{align}
\Sigma_{\text{mix}}^{ij} &=4\text{Re}\int_{0}^\infty \frac{(\partial_k \hat{h}_m(f) \partial_l \hat{h}^{\star}_m(f)) (S_{\text{conf}}(f)+S_{\text{n}}(f))}{S_{\text{n}}^2(f)}.
\end{align}

In contrast to this, the standard approach when modelling the confusion background is to combine the instrumental and confusion noises into a single noise term, $N = n + \Delta  H_{\text{conf}}$. Then the standard parameter estimation formalism can be used, with the substitution $S_{\text{n}} (f) \rightarrow S_{\text{n}}(f) + S_{\text{conf}}(f)$ in the inner product~\eqref{eq:inn_prod}. In this case the inference uncertainties are given by the inverse of the Fisher matrix, $\Gamma^{-1}_{\text{n}+\text{conf}}$, where
\begin{equation}
 \Gamma_{\text{n}+\text{conf}}^{ij} =4\text{Re}\int_{0}^\infty \frac{(\partial_k \hat{h}_m(f) \partial_l \hat{h}^{\star}_m(f))}{S_{\text{n}}(f)+S_{\text{conf}}(f)}.
 \label{eq:confvar}
\end{equation}
The variance given by Eq.~\eqref{eq:biasconfvar} is, in general, larger than that predicted by Eq.~\eqref{eq:confvar}. This is because it has been derived by maximizing the standard likelihood as an estimator of the parameters, which is no longer the correct likelihood when random confusion noise is included in the  model. Expression~\eqref{eq:confvar} gives the variance of the true maximum likelihood estimator, which is known to be the minimum variance unbiased estimator and must therefore be smaller than~\eqref{eq:biasconfvar}. Incorporating the confusion noise uncertainty into the PSD is the correct thing to do when Eq.~\eqref{eq:confPSD} is known to be a good approximation, but the formalism described here can be used when that equation is not valid, and to assess when confusion noise is likely to be problematic for parameter estimation. As a final remark, we note that in the limit that there are a large number of sources contributing to the confusion background, the central limit theorem allows us to approximate the probability distribution of the parameter bias correction, $p(\Delta  \boldsymbol\theta_{\text{conf}})$, as a Gaussian with mean $\boldsymbol{\mu}_{\text{conf}}$, given by Eq.~\eqref{eq:mu_conf}, and covariance $\Sigma_{\text{conf}}$. The correct statistical procedure of marginalising the likelihood for $d(t)-\Delta H(t)$ over the confusion noise distribution thus amounts, in the linear signal approximation, to shifting the mean by $\boldsymbol{\mu}_{\text{conf}}$ and adding $\Sigma_{\text{conf}}$ to the covariance. The results described here can therefore be used not only to assess when confusion is important but also to compute leading order corrections to posterior parameter estimates arising from the presence of confusion. 

\section{Numerical Routines}
\label{app:num_routines}
In this appendix, we provide more details on how we sample our signals in the frequency domain. We begin by choosing a starting frequency $f_{0}$ and final frequency determined by the last stable orbit in a Schwarzschild spacetime $f_{\text{max}}= c^{3}/6\sqrt{6}\pi GM$. The calculated time to merger is then predicted through the 3.5PN chirp time (see Eq.(3.5a) of~\citep{Allen:2005fk}). Invoking Shannon's sampling theorem \citep{shannon1949communication}, the spacing between time points $\Delta t$ is chosen to be $\Delta t = 1/(2f_{\text{max}})$. For multiple signals, we choose the minimum sampling interval common to all waveforms for given mass parameters. In doing so, we find the length of the signal $N_{t} = \lfloor t_{obs}/\Delta t \rfloor$ in the time domain. Combining all these elements, one is able to construct a list of sampling frequencies $f = [0,\Delta f, 2\Delta f, \dots, \lfloor (N_{t} - 1)/2\rfloor\Delta f]$ for $\Delta f = 1/N_{t}\Delta t$. Given the discrete Fourier frequencies, it is then possible to construct waveforms using \eqref{eq:signal_model_SPA}.

Noise is generated in the frequency domain with real and imaginary parts drawn separately from Gaussian distributions with equal variance and zero mean. Discretising equation \eqref{eq:Wiener-Khinchin-Theorem_freq}, it's easy to show that the variance of both real and imaginary parts are equivalent to 
\begin{equation}
    \sigma^{2}(f_{i}) = N_{t}S_{n}(f_{i})/4\Delta t.
\end{equation}

Finally, in order to calculate various quantities involving inner products (Fisher matrices, SNRs and likelihoods), we use the discrete analogue of \eqref{eq:inn_prod},

\begin{equation}\label{eq:discrete_inn_prod}
    (a|b) \approx 4\Delta f \ \text{Re} \sum_{i=0}^{\big \lfloor \frac{N_{t}-1}{2} \big 
    \rfloor }\frac{\hat{a}(f_{i})\hat{b}^{\star}(f_{i})}{S_{n}(f_{i})}.
\end{equation}

\section{Fisher Matrices and their validation}\label{app:Fisher_Matrix}

The Fisher Matrix \eqref{eq:Fish_Matrix} can be calculated through inner products of waveform derivatives. We choose to use a second order finite difference method, 
\begin{equation}
    \frac{\partial h_m(f;\Theta^i)}{\partial \Theta^i} \approx \frac{h_m(f;\Theta^i+ \delta\Theta^i)-h_m(f;\Theta^i -\delta\Theta^i)}{2 \delta\Theta^i}
\end{equation}
Fisher matrices in gravitational wave astronomy have high condition numbers, which influence our ability to obtain reliable parameter precision estimates. We invert our Fisher matrices using the high precision arithmetic Python package \texttt{mpmath} \citep{mpmath}. This was done in order to mitigate instabilities arising from computing the inverse of the potentially badly conditioned matrix ~\citep{wen2005detecting,abuse_fisher,porter2009overview,rodriguez2012verifying,gair2013testing,porter2015fisher,amaro2018relativistic,PhysRevD.102.124054}. A criterion to establish the stability of the inverse Fisher matrix based on the (1-norm) absolute value reads $\big|\Gamma^{-1}\Gamma - \texttt{I}\big|_{1} \leq 10^{-3}$, where $\texttt{I}$ is the identity matrix~\citep{Gupta:2020lxa}. We used $\sim$ 500 decimal digits and found  $\Gamma^{-1}\Gamma = \texttt{I} - \epsilon_{ij} \texttt{I}$ with $\max_{i,j}\{|\epsilon_{ij}|\} \approx 10^{-14}$, even with condition numbers $\sim 10^{21}$. This gives us confidence that the numerical inversion of our Fisher matrix is both numerically robust and accurate. 

To validate our results, we carry out a Markov-Chain Monte-Carlo (MCMC) with the goal to match our Fisher matrix results in a high-SNR regime. Our Bayesian analyses are carried out using \texttt{emcee} \citep{ForemanMackey:2012ig} and an appropriate modification of the code developed in~\citep{PhysRevD.102.124054}. The posteriors are sampled with \texttt{emcee} using a Whittle log-likelihood~\eqref{eq:whittle_likelihood} and flat priors. A publicly available implementation of the MCMC illustrations carried out with \texttt{emcee} can be found at \url{https://github.com/aantonelli94/GWOP}.
The latter code is based on a standard Metropolis-Hastings algorithm~\citep{metropolis1953equation}. A publicly available implementation can be found at \url{https://github.com/OllieBurke/Noisy_Neighbours}.  For this algorithm, we chose a proposal distribution equivalent to a multivariate Gaussian with covariance matrix equal to a scaled variate inverse of the Fisher Matrix. By pre-multiplying the inverse Fisher matrix $\Gamma^{-1}$ by $N_{\text{sources}}$, we found better acceptance ratios $\sim 30\%$ [near the optimal acceptance rate for non-single parameter studies~\citep{roberts1997weak}].

\section{Predicting waveform and confusion noise biases with ET}
\label{app:ET_gf}

\begin{figure*}
    \centering
    \includegraphics[height = 5.5cm, width = 17cm]{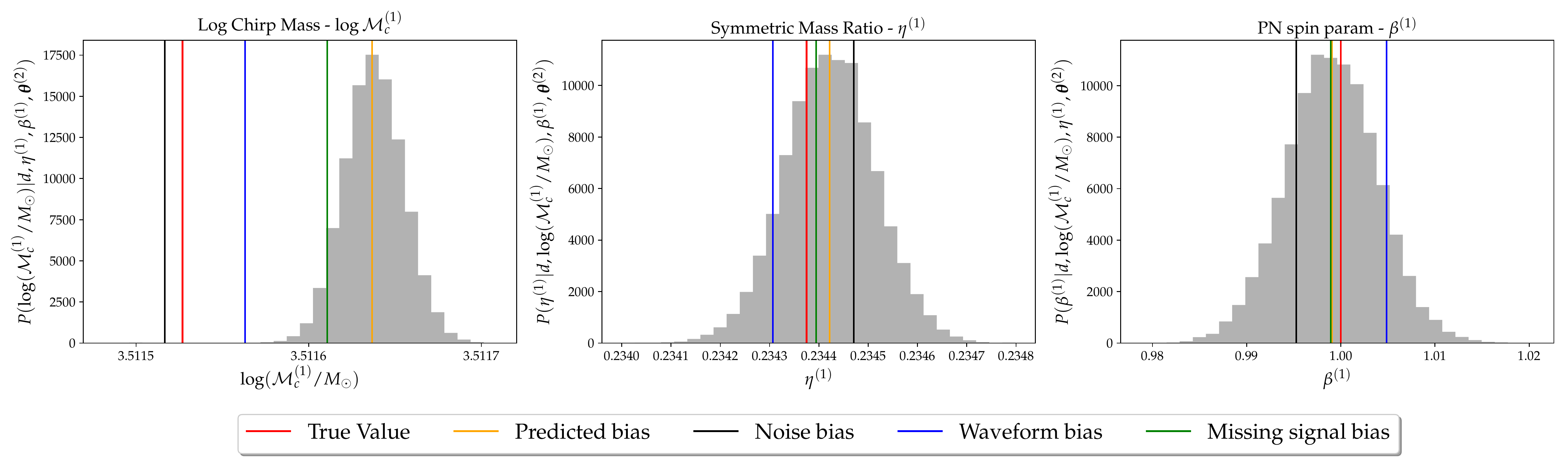}
    \caption{Same as Figure \ref{fig:posteriors_LISA} but for the ET configuration of Appendix~\ref{app:ET_gf}.}
    \label{fig:posteriors_LIGO}
\end{figure*}

In this appendix, we repeat the analysis of Sec.~\ref{sec:results} for a source in ET.
We use the same data stream as~\eqref{eq:global_eq}, modelling $N_\text{fit}=2$ simultaneously-fitted signals in a similar manner. We pick waveform errors $\epsilon = 0.02$ and a starting frequency $f_{0} = 5$Hz. As for confusion noise, we construct it with a series of missed signals which we model without errors. We report the parameters for both fitted and missed sources in Tab.~\eqref{tab:sec_app_table}. The SNRs of the fitted signals are $\mathcal{O}(10^{3})$, those of the missed signals $\lesssim 1000$ (with the lowest $\sim 200$). The SNRs of the missed signals for ET are noticeably high, and would likely be detected in a future analysis. However, for sake of example, treat these signals as missed signals in the parameter estimation scheme. The predictions for the biases of all parameters, Fig.~\ref{app:ET_gf}, show that the formalism can predict the mean of the posterior as remarkably well as in the case of LISA. The individual bias contributions, Fig.~\ref{fig:posteriors_LIGO}, confirm that biases can deconstructively interfere.

\begin{table}
	\centering
	\caption{Parameter configurations for the ET case.}
	\label{tab:sec_app_table}
\begin{tabular}{c|cccc}
\multicolumn{1}{l|}{}          & \multicolumn{4}{c|}{Fitted}  \\ \hline
$i$                    & $M/M_{\odot}$ & $\eta$ & $\beta$ & $D_{\text{eff}}/\text{Mpc}$  \\ \hline
1                      &   80             & 0.234       &  1    & 400    \\
2                      &   70             & 0.204       &  5    & 40     \\
\multicolumn{1}{l|}{}          & \multicolumn{4}{c|}{Missed}   \\ \hline
$i$                    & $M/M_{\odot}$ & $\eta$ & $\beta$ & $D_{\text{eff}}/\text{Mpc}$  \\ \hline
1                      & 2.22  & 2.708 & 5.04  & 259.93   \\
2                      & 2.886 & 0.247 & 3.882 & 253.36   \\
3                      & 4.395 & 0.2264 & 5.539 & 324.227    \\
4                      & 6.452 & 0.1991 & 4.404 & 305.828   
\end{tabular}
\end{table}

\begin{figure*}
     \centering
     \includegraphics[height = 15cm,width = 18cm]{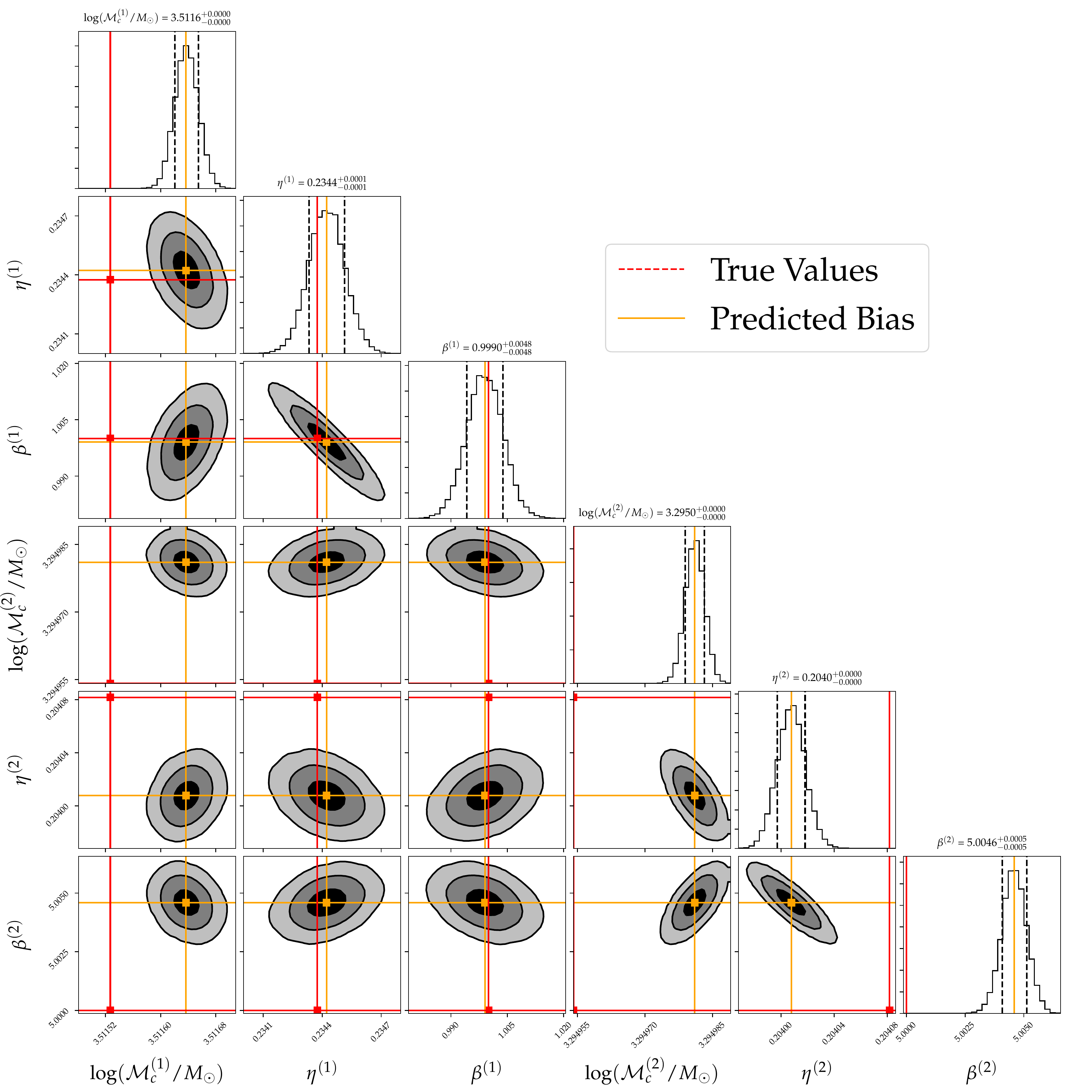}
     \caption{Same as Fig.~\ref{fig:LISA_Corner} but for the ET configuration of Appendix~\ref{app:ET_gf}.}
 \end{figure*}

\end{document}